


%
%

\def\logt{$\log T_{\rm eff}$}
\def\logg{$\log g$}
\def\teff{$T_{\rm eff}$}
\def\lt{\log T_{\rm eff}}
\def\lg{\log g}
\def\moverdot{{^{\rm m}\!\!\! .\,}}
\def\logl{$\log L/{{\rm L}_\odot}$}
\def\loglum{\log L/{{\rm L}_\odot}}
\def\mdot{$\dot M$}
\def\vterm{$v_\infty$}
\def\bminu{$[B-U]$}
\def\bminl{$[B-L]$}
\def\kms{\, {\rm km}\, {\rm s}^{-1}}
\def\ergps{\, {\rm ergs}\, {\rm s}^{-1}}
\def\pc{\, {\rm pc}}
\def\cc{\, {\rm cm}^{-3}}
\def\yr{\, {\rm yr}}
\def\Myr{\, {\rm Myr}}
\def\magn{^{\rm m}}
\def\um{\, \mu {\rm m}}
\def\av{$A_V$}
\def\ea{{\rm et~al.}}
\def\Msol{\, {\rm M}_\odot}
\def\HI{H{\sc i}\ }
\def\HII{H{\sc ii}\ }
\def\mv{m_{\scriptscriptstyle V}}
\def\mb{m_{\scriptscriptstyle B}}
\def\vbluw{$V\! BLUW$\ }
\def\deg#1{\ifmmode{#1{^\circ}}\else{#1${^\circ}$}\fi}
%
%
\def\aj#1{AJ\ #1}
\def\apj#1{ApJ\ #1}
\def\aap#1{A\&A\ #1}
\def\araa#1{ARA\&A\ #1}
\def\aapss#1{A\&AS\ #1}
\def\apjss#1{ApJS\ #1}
\def\apss#1{Ap\&SS\ #1}
\def\mn#1{MNRAS\ #1}

\def\pasj#1{PASJ\ #1}


  \MAINTITLE={ The Orion OB1 association
		\FOOTNOTE{ Based on \vbluw Photometry
		obtained with the 91-cm Dutch Telescope at ESO, La
		Silla } }

  \SUBTITLE={ I. Stellar content }

  \AUTHOR={ A.G.A. Brown@1, E.J. de Geus@{2,3} and P.T. de Zeeuw@1 }

    \OFFPRINTS={ A.G.A. Brown }

  \INSTITUTE={ @1 Sterrewacht Leiden, P.O.~Box 9513, 2300 RA, Leiden,
		  The Netherlands
	       @2 Astronomy Department, Caltech, Pasadena, CA 91125,
		  USA
	       @3 Astronomy Department, University of Maryland,
		  College Park, MD 20742, USA }

  \DATE={ Received\dots, accepted\dots }

  \ABSTRACT={ Walraven photometry of established and probable members
of the Orion OB1 association is presented. Effective temperature,
surface gravity, luminosity and mass are derived for all stars, using
atmosphere models by Kurucz (1979).  Absolute magnitudes are
calculated using the Strai\v zys and Kuriliene (1981) tables. Distance
moduli and visual extinctions are determined. A comparison of the
visual extinctions to IRAS $100\um$ data shows that the near edge of
the Orion A and B clouds lies at a distance of $\sim 320\pc$, while
the far edge is at $\sim 500\pc$.  A method for deriving the ages of
the subgroups by comparing theoretical isochrones to the observations
in the \logg , \logt\ plane is presented. The derived ages suggest,
contrary to earlier studies, that subgroup 1b is younger than 1c,
which can possibly be explained by past geometries of the system of
stars and gas. The initial mass function for Orion OB1 is derived with
the aid of the Kolmogorov-Smirnoff test. Through extensive
simulations, we show that it is very difficult to derive accurately
the IMF from the available data. To within somewhat weak limits the
IMF is found to be of the form $\xi (\log M)=AM^{-1.7\pm 0.2}$ for all
subgroups. The energy output of the subgroups in the form of stellar
winds and supernovae is calculated and compared to the observed size
and expansion velocity of the Orion-Eridanus bubble. It is shown that
the energy output of the association can account for the morphology
and kinematics of the ISM. }

   \KEYWORDS={ stars: fundamental parameters --
Hertz\-sprung-Russell diagram -- open clusters and associations: general,
individual: Orion OB1 -- ISM: bubbles }

  \THESAURUS={ 08.06.3; 08.08.1; 10.15.1; 10.15.2 Orion OB1; 09.02.1 }

\maketitle

\titlea {Introduction}
OB associations play an important role in the large scale propagation
of star formation and in the shaping and evolution of the interstellar
medium. The OB star complex in Orion is an excellent laboratory for an
investigation of these processes. It contains the four subgroups of
the Orion OB1 association as well as the Orion Molecular Cloud (OMC)
which is one of the nearest sites of active OB star formation. The
subgroups were identified by Blaauw (1964), and differ in age and in
content of gas and dust. The different evolutionary phases of the
subgroups make the Orion region well suited for studying the influence
of OB associations on the surrounding interstellar medium and for
constraining scenarios of sequential star formation, such as suggested
by Elmegreen \& Lada (1977). Knowledge of the stellar content also
provides clues to the origin of the initial mass function.

The stars in Orion OB1 have been studied by a number of authors (see
Blaauw 1964; Genzel \& Stutzki 1989). Early photometric and
spectroscopic work by Parenago (1954) and Sharpless (1952, 1962) was
followed by the massive investigation by Warren \& Hesser (1977, 1978,
hereafter WH) based on $uvby\beta$ photometry. These authors derived
distances and ages for the subgroups. Proper motion investigations
were carried out by Parenago (1953), Strand (1958), Lesh (1968) and
more recently by Van Altena \ea\ (1988), Jones \& Walker (1988) and
Smart (1993).

The interstellar medium near Orion OB1 contains several large scale
features, including H$\alpha$ emission extending to Eridanus, partly
observable as Barnard's loop, and a hole in the \HI distribution,
which is surrounded by expanding shells (Goudis 1982).  Reynolds \&
Ogden (1979) and Cowie \ea\ (1979) argued that the coherent gas
motions in Orion are the result of a series of supernova events which
took place up to $4\times 10^6$ yr ago, but they ignored the effects
of stellar winds. In the past fifteen years a wealth of new data has
been gathered on the large scale interstellar medium in Orion, through
surveys in $^{12}$CO (Maddalena \ea\ 1986), $^{13}$CO (Bally \ea\
1987), CS (Lada \ea\ 1991), the far-infrared (IRAS sky survey) and \HI
(Chromey \ea\ 1989; Green 1991; Green
\& Padman 1993; Hartmann \& Burton 1993). At the same time, the theory of
stellar winds has been developed to the extent that their impact on
the surrounding medium can now be readily estimated (Kudritzki \ea\
1989; McCray
\& Kafatos 1987; de Geus 1991, 1992). A new investigation of the stellar
content of the Orion OB1 association, especially in relation to the
large scale structures in the surrounding ISM, is therefore justified
and timely.

This paper is organized as follows. The next section describes the
sample of stars and the photometric observations. In section 3 the
method used for deriving physical parameters of the stars (\logg\ and
\logt) from \vbluw photometry is summarized. Section 4 describes a
number of properties of the stellar content of the Orion OB1
association. These include ages and distances of the subgroups,
derivation of membership criteria, correlations between the visual
extinctions of the stars and gas and dust around the association
(using IRAS skyflux data) and a calculation of mass functions and the
energy output of the subgroups in Orion OB1. The relation between the
stars and the large scale structures in the surrounding ISM is
described in section 5. Section 6 contains the conclusions and
suggestions for future work.

\titlea {Observations}
The stars studied in this paper are established or probable members of
the Orion OB1 association.

In order to obtain a list of potential members down to spectral type
F, the CSI catalogue (Jung \& Bischoff 1971) was searched with a
number of selection criteria. The coordinates were limited by: $
196^\circ \leq\ell \leq 217^\circ$, and $ -27^\circ \leq b \leq
-12^\circ$ (Blaauw 1964).  All O and B stars in this area were
selected, while A and F stars were required to have $\mb \geq 7\magn$,
respectively $\mb \geq 9\magn$, to avoid a large number of foreground
objects. The CSI catalogue is complete down to $\mv \approx 9\magn$,
and is limited at $\mv \approx 12\magn$. This selection procedure
resulted in 1318 stars, which were included in a 1982 HIPPARCOS
proposal to observe all OB associations within $800\pc$ from the Sun.

Photometric observations were carried out between 1982 and 1989 with
the 91-cm Dutch telescope at ESO, in the \vbluw Walraven system (Lub
\& Pel 1977; de Geus, de Zeeuw \& Lub 1989, hereafter GZL). This
provided measurements for 986 of the 1318 stars, tabulated by de Geus
\ea\ (1990). These stars are the subject of this paper. The mean rms
errors in the observed intensities are: $\bar\sigma (V)=0.0016$,
$\bar\sigma (V-B)=0.0010$, $\bar\sigma (B-U)=0.0012$, $\bar\sigma
(U-W)=0.0015$, $\bar\sigma (B-L)=0.0012$. Multiplication by 2.5
provides the mean rms errors in magnitudes. The stars were divided in
two categories: 373 priority 1 stars which are a) stars of spectral
type O and B, and/or b) stars that are established or probable members
of the association according to WH, who used proper motion data,
radial velocities and $U\! BV$ as well as $uvby\beta$ photometric
observations. The remaining 613 stars have priority 2.

The final HIPPARCOS Input Catalog contains 699 stars of our original
sample of 1318 stars in Orion OB1, of which 236 are priority 1, and
463 are priority 2.

\titlea {Derivation of physical parameters}
The derivation of physical parameters from photometric data involves a
number of steps in which theoretical and empirical transformations are
used. We follow the procedure used by GZL in their study of Sco OB2.

The first step is the calculation of reddening-independent colours. In
the Walraven system these are defined as (Lub \& Pel 1977):
$$\eqalign{
[B-U] &=(B-U)-0.61(V-B)\; , \cr
[U-W] &=(U-W)-0.45(V-B)\; , \cr
[B-L] &=(B-L)-0.39(V-B)\; . \cr}                                 \eqno(1)$$
The colour $(V-B)$ essentially measures the reddening, $[B-U]$ is an
indicator of \logt, and for O and B stars $[U-W]$ and $[B-L]$ each
depend mostly on
\logg. The errors in the reddening-independent colours are caused by
observational errors and by possible local variations in the value $R$ of
total to selective extinction. WH find that a normal reddening law is
appropriate for Orion OB1, except for selected regions such as the Orion
Nebula. We have therefore adopted the conventional value $R = 3.2$. This is
reflected in the coefficients of Eq.\ (1).

\begfig 8cm
\figure {1} {
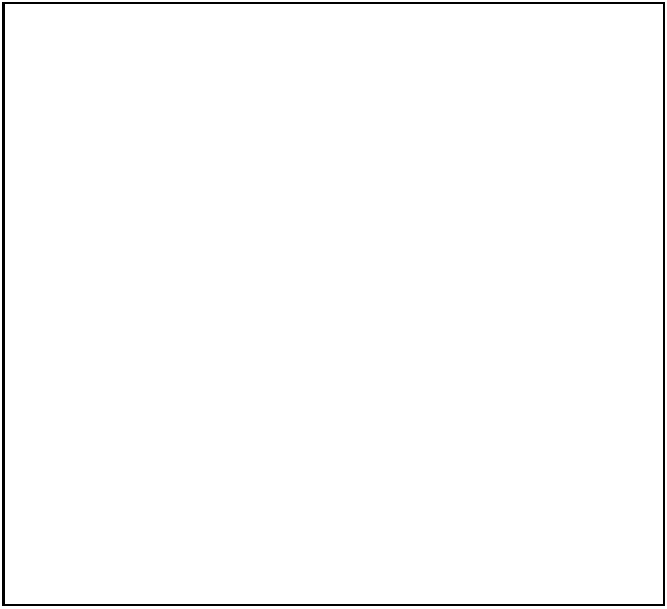
The Kurucz grid in the reddening-independent two-colour diagram of
\bminu\ vs.~\bminl . Solid lines indicate constant \logt\ and dashed
lines constant
\logg. Values of \logt\ and \logg\ are indicated.  Points are the observations
for our sample of stars in Orion
}
\endfig

The next step is the determination of effective temperature and
surface gravity. These were derived from the
red\-de\-ning-inde\-pen\-dent colours by employing a grid of
theoretical colours for a wide range of \teff\ and \logg, obtained by
convolution of the Kurucz (1979) atmosphere models with the Walraven
passbands. The availability of three reddening-independent colours
allows the construction of two independent reddening-free
colour-diagrams; \bminu\ vs.\ $[U-W]$ and \bminu\ vs.\ \bminl. Due to
uncertainties in the calibration of the Kurucz grid in the \bminu\
vs.\ $[U-W]$ plane (Brand \& Wouterloot 1988), only the $[B-U]$ vs.\
$[B-L]$ diagram was used. Figure 1 shows the Kurucz grid in this
diagram together with all our programme stars.  The values of \logt\
and \logg\ were determined by two-dimensional linear interpolation in
the grid. The Kurucz models are unreliable for stars with $T_{\rm eff}
< 8000$K, and we therefore excluded all stars with $[B-U]<0.3$ and
$[B-L]>0.13$. Stars with $\lg\ge 4.5$ just outside the grid were
assigned \logg\ and \logt\ by taking into account their observational
errors, or by using an extrapolation of the grid as described by Brand
\& Wouterloot (1988).  Other stars outside the grid were excluded. A
number of stars lie in the part of the diagram where the Kurucz grid
folds over itself, resulting in multiple solutions. These were checked
carefully by investigating the extinctions (which have to be positive)
and by comparing temperatures and spectral types. Usually only one
solution remained and \logg\ and \logt\ could be assigned. The errors
in the derived \logt\ and \logg\ are due to the propagation of errors
in the original input-colours and to systematic errors in the Kurucz
grid. The errors in the grid are largest for temperatures above
$25000\, {\rm K}$ (and below $8000\, {\rm K}$). This results in
$\bar\sigma (\lt)\approx 0.015$ for $\lt <4.3$ and $\bar\sigma
(\lt)\approx 0.03$ for $\lt\ge 4.3$. The errors in \logg\ are $\approx
0.1$ and $\approx 0.25$, respectively.

The absolute visual magnitude $M_V$ and the absolute bolometric
magnitude $M_{\rm bol}$ were derived from \logt\ and \logg\ by means
of the calibration Tables of Strai\v zys \& Kuriliene (1981). These
give \logt, \logg, $M_V$, $M_{\rm bol}$, spectral type and luminosity
class, based on a combination of empirical calibrations and
theoretical calculations. We employed two-dimensional linear
interpolation in these Tables to derive $M_V$ and $M_{\rm bol}$. The
luminosity was calculated from the bolometric magnitude by:
$$\log {L\over {{\rm L}_\odot}}=-0.4M_{\rm bol}+1.888\; .
\eqno(2)$$

As can be seen in Fig.~2 of GZL, the Kurucz grid and the Strai\v zys
\& Kuriliene calibrations do not overlap entirely. This implies that a
number of stars have \logg, \logt\ solutions from the Kurucz grid, but
can not be assigned a magnitude. For stars on the main-sequence, but
with \logg\ values just below the ZAMS line in Fig.~2 of GZL, we used
\logt\ to assign a photometric spectral type. The luminosity class was
determined by assuming that stars with $T_{\rm eff}\ge 25000\, K$ are
of luminosity class {\sc v}. The other main-sequence stars were
assigned luminosities corresponding to a ZAMS star in the Tables of
Strai\v zys \& Kuriliene. The absolute magnitude of a star is very
sensitive to both \teff\ and \logg, so that the errors in derived
temperature and gravity cause large uncertainties in the values of
$M_V$ and $M_{\rm bol}$. We estimate $\bar\sigma (M_{\rm bol})\approx
0\moverdot 3$.

The Kurucz grid can also be used in observed-colour space to transform
\logt\ and \logg\ to the intrinsic colours. The colour excess in the
Walraven system, $E_{(V-B)}$ is the difference of the observed colour
$(V-B)$ and the intrinsic colour $(V-B)_0$:
$E_{(V-B)}=(V-B)-(V-B)_0$. The colour excess in the Johnson system,
$E_{(B-V)}$, follows from that in the Walraven system by (GZL):
$$E_{(B-V)}=2.39E_{(V-B)}-0.17E^2_{(V-B)}\; .               \eqno(3)$$
The total visual extinction $A_V$ is then given by $A_V=3.2 E_{(B-V)}$.

To determine the distance modulus of a star we need the apparent
visual magnitude $\mv $ in the Johnson system, because the Strai\v zys
\& Kuriliene calibrations give the absolute magnitude in this
system. The following transformation formula was used to derive $\mv$
from the Walraven $V$ and $(V-B)$ (GZL):
$$\mv =6.886-2.5V-0.082(V-B)\; .         \eqno(4)$$
The distance modulus then follows from the usual relation: $5\log
D-5=\mv -M_V-A_V$. The mean errors in $A_V$ and $\mv$ are $0\moverdot
03$ and $0\moverdot 015$, respectively. The error in the distance
modulus is dominated by the error in $M_V$, and is approximately
$0\moverdot 3$.

The results are presented in Table 1. The first three columns give the
identification of each star: column 1 is the HD or BD number, column 2
lists the number of the star in the HIPPARCOS Input Catalogue (HIC,
Turon \ea\ 1992), and column 3 gives the name. Columns 4--6 list \logt
, \logg\ and $\log L/L_\odot$. Column 7 gives $\mv$. Column 8 lists
the absolute magnitude and column 9 the visual extinction.  Column 10
lists the distance modulus and column 11 gives an indication of the
membership: members according to the criteria described in Section 4
have a blank, possible members based on the findings from
pho\-to\-me\-try are designated with PM and non-members are denoted by
NM. An asterisk in column 11 means that the star is included in Fig.\
6a and a plus means that it is included in Fig.\ 6b (see section
4.3). Column 12 gives the spectral type according to the
MK-classification, taken from the HIPPARCOS Input Catalogue. De Geus
\ea\ (1990) give the original photometric data, the positions on the
sky and the priority (1 or 2) for all the stars.

Absolute magnitudes and distance moduli for the stars HD 37022, HD
37128 and HD 38771 could not be determined because the stars lie
outside the Kurucz grid or \logg\ and \logt\ are outside the range of
the Strai\v zys \& Kuriliene calibrations. Physical parameters for HD
37128 and HD 38771 were taken from Lamers \& Leitherer (1993) and
Vilkoviskij \& Tambovtseva (1992), respectively.

\titlea {Properties of the Orion OB1 association} Blaauw (1964)
divided Orion OB1 into four subgroups: 1a, which contains the stars to
the northwest of the Belt stars; 1b, containing the group of stars
located around the Belt (including the Belt stars themselves); 1c, in
which the stars around the Sword are included; and 1d, which contains
the stars in and close to the Orion Nebula (including the Trapezium
stars). Subgroups 1b, 1c and 1d were subdivided further in subsequent
studies. WH split subgroup 1b into three parts because Hardie \ea\
(1964) and Crawford \& Barnes (1966) found that the distance of the
Belt stars increases from west to east. We also used this subdivision
in the determination of membership criteria (Section 4.1), but we
found no significant differences in their mean distances, and no trend
with right ascension for the 1b stars as claimed by WH. So for all
other purposes subgroup 1b was treated as a whole. Morgan \& Lod\' en
(1966) and Walker (1969) divided 1c into several smaller subgroupings
located close to the Orion Nebula.  WH found no evolutionary
differences between these groups, so we decided to treat subgroup 1c
as a whole also. WH divided subgroup 1d in an outer Nebula region and
an inner Nebula region, in the hope that an accurate distance modulus
could be determined for the outer region which could then be applied
to the inner region. We used the same subdivision for the distance
determination.

\titleb {Distances to the subgroups; photometric membership
determination} Ideally one would use the established proper motion
(and radial velocity) members of a stellar group to derive its mean
distance and distance spread, and then use these to determine
photometric membership for stars without measured proper motions or
radial velocities. However, we would like to point out here that this
procedure works well for open clusters, but that in the case of
associations assigning membership based on the motions of the stars is
a more complicated problem. This is due to the fact that associations
are formed in a much larger region of space (of the order of tens of
parsecs, the size of GMCs) and with much lower stellar space densities
than open clusters. As a consequence, associations are very loose
stellar aggregates in which each star is individually subject to
Galactic forces. Therefore they will not stand out clearly on the sky
(making it difficult to assign spatial boundaries) or in proper motion
diagrams. Furthermore, stars formed in different regions of the
parental cloud might have different motions from the beginning, even
if they formed at the same time. For a star at the periphery of the
star forming region it will also be difficult to assign membership
based on its motion: early on it might have acquired an anomalous
motion as a consequence of Galactic forces. Stars in associations
might also acquire anomalous motions as a consequence of supernova
explosions in binaries (possibly leading to the formation of runaway
stars, see e.g., Blaauw 1961) or because of dynamical ejection
mechanisms (Gies \& Bolton 1986, Leonard \& Duncan 1988, 1990). There
is also the possibility of non--coeval star formation within a
subgroup, which can result in stars of different ages moving
differently. In short, due to the spread in space and possibly in time
of the formation of an association, the term "established proper
motion member" (as derived from earlier studies) has less meaning than
it has in the classical proper motion studies of open clusters. One
should find better criteria (apart from, e.g., converging proper
motion vectors), taking into account the formation history, for
assigning membership in associations.

Nevertheless, we decided to use a subset of the stars for which WH
derived membership based on proper motions, radial velocities and
photometry and regard these as the established members of Orion OB1.
These are the priority 1 stars in our programme and we used them to
derive mean distance moduli and spreads. We used only the B stars, for
which the Kurucz grid is reliable. We applied $3\sigma$-clipping to
derive the mean distance modulus, with $\sigma$ being the rms spread
around the mean. The results are listed in Table 2 together with the
distance moduli found by WH and, for both studies, the number of stars
used to derive the results. Our distance moduli are systematically
smaller than those of WH, because our zero age main-sequence absolute
magnitude calibration is systematically fainter than the $M_V(\beta)$
calibration used by WH. Anthony-Twarog (1982) reanalyzed the data in
WH with the aid of a revised calibration of the $uvby\beta$ system
(Crawford 1978). She finds distance moduli for the subgroups that are
consistent with our results to within $0\moverdot 1$.  The mean
distance moduli of the three subdivisions of 1b are the same within
the errors. The distance to subgroup 1d is very uncertain because it
is difficult to obtain accurate photometry in this highly nebulous
area (the small rms spread in distance modulus is not significant,
only three stars were included in the calculation of $\langle {\rm
dm}\rangle$).

\begtabfull
\tabcap {2} {Mean distance moduli of the subgroups and rms spreads}
\halign{#\hfil&&\quad#\hfil\cr
\noalign{\hrule\medskip}
\omit&\multispan3\hfil WH\hfil&\multispan3\hfil \vbluw\hfil\cr
\noalign{\medskip\hrule\medskip}
Subgroup&$\langle {\rm dm}\rangle$&$\sigma$&$n$&$\langle {\rm
dm}\rangle$&$\sigma$&$n$\cr
\omit&(mag)&(mag)&\omit&(mag)&(mag)&\omit\cr
\noalign{\medskip\hrule\medskip}
	a&8.0&0.46&60&7.9&0.52&53\cr
	b1&8.5&0.53&14&7.8&0.39&8\cr
	b2&8.2&0.29&24&7.8&0.45&17\cr
	b3&8.0&0.55&14&7.8&0.46&6\cr
	c&8.2&0.49&78&8.0&0.49&34\cr
	d&8.4&0.53&6&7.9&0.25&3\cr
\noalign{\medskip\hrule}}
\endtab

WH based their membership assignments on proper motions, radial
velocities and distance moduli. We carefully checked all the priority
1 stars with distances outside our $2\sigma$ limits. A number of these
stars were marked as possible non-members by WH. While examining these
stars, we took into account errors in the distance modulus, undetected
duplicity, spectral peculiarities, nebulosity influence on the
photometry and evolutionary effects. In this way we could exclude
membership for a number of stars marked as doubtful members by
WH. Overall we find good agreement between our membership list and
that of WH.

\begfig 8cm
\figure {2}{
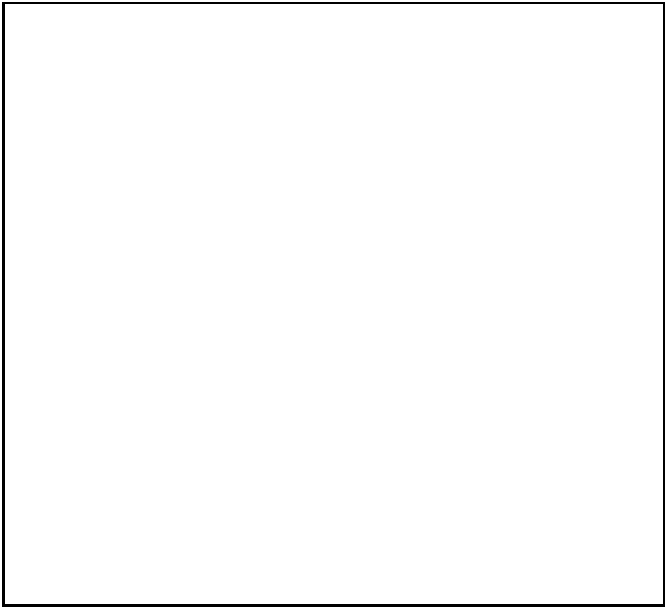
Distance modulus vs.\ right ascension for the stars that are members
of subgroup 1b
}
\endfig

\begfig 8cm
\figure {3}{
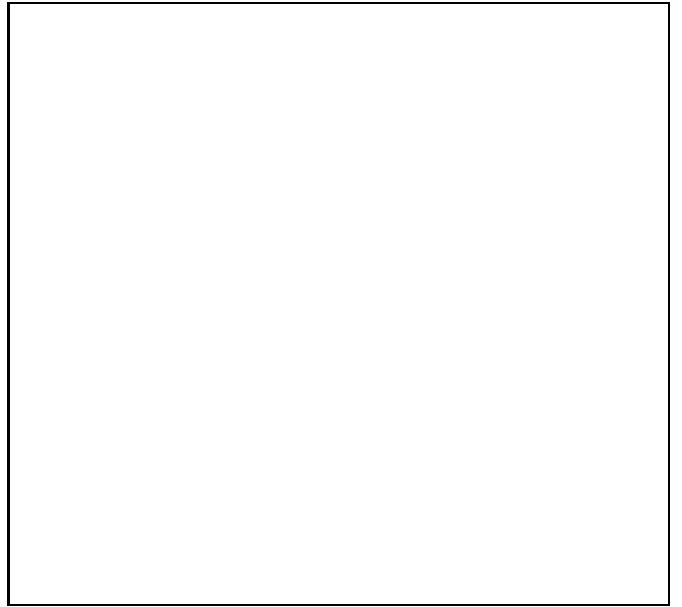
Reddening-independent two-colour diagram for all program stars in
Orion OB1. The solid line is the theoretical zero-age main-sequence
}
\endfig

OB associations generally have linear sizes between 10 and $100\pc$
(Blaauw 1991). The angular size of the Orion OB1 subgroups corresponds
to sizes in the range of 20 to $45\pc$, and it is likely that $100\pc$
is a safe upper limit to the depth of the association along the line
of sight. At the distance of Orion OB1 ($> 350\pc$) this corresponds
to $0\moverdot 3$ intrinsic spread in the distance moduli. This means
that the observational errors in the individual distance moduli are
the dominant source of the spread around the mean, and that the extent
of the subgroups along the line of sight cannot be resolved. The
membership assignments for the priority 2 stars were therefore based
on the following criteria: if the distance modulus of a star is within
$2\sigma$ from the mean it is considered a member. If the star has a
distance modulus that is between $2\sigma$ and $3\sigma$ from the mean
it is considered a possible photometric member. Otherwise it is a
non-member.

The priority 2 stars in subgroups 1a and 1c cover a much larger area
on the sky than the priority 1 stars in these groups. This means that
selecting members only on basis of their distance modulus is likely to
introduce a number of field stars in the list of members. This can
never really be avoided in photometric membership determinations. For
subgroup 1d we had too few stars in the outer Nebula subdivision to
calculate a reliable mean distance modulus and spread.  Instead we
used the proper motion study of the Orion Nebula cluster (the
Trapezium region) by Van Altena \ea\ (1988) to assign membership for
all the programme stars in 1d.

In order to make our list of association members as reliable as
possible we also used the proper motion data from the PPM catalogue
(R\" oser \& Bastian 1989; Bastian \ea\ 1991). Vector point diagrams
(VPD) were constructed for the subgroups and clear outliers were
excluded as members. WH obtained the following mean centennial proper
motions for all subgroups: $\langle\mu_\alpha\rangle =-0^{\rm s}\!\!
.03\pm 0^{\rm s}\!\! .03$ and $\langle\mu_\delta\rangle =-0^{''}\!\!\!
.1\pm 0^{''}\!\!\! .5$. The mean motions found by WH are consistent
with the mean motions we find for our member stars from the PPM
data. The outliers in our VPD on average lie at more than $\sim
0^{''}\!\!\!  .02\, {\rm yr}^{-1}$ from the centroid of the subgroup,
well outside the $3\sigma$ limits on the mean value of the proper
motion. The errors in the PPM proper motions are $\sim 0^{''}\!\!\!
.003\, {\rm yr}^{-1}$. It turns out that almost all of the discarded
stars are on the near side of the association between the $2\sigma$
and $3\sigma$ limits on the mean distance moduli.

As can be seen in Table 2 WH found that the distance moduli of stars
in subgroup 1b increase from west to east. Figure 2 shows our distance
moduli for the members of this subgroup as a function of right
ascension. No trend is evident. The correlation coefficient between
distance modulus and RA is $0.03$ (at a significance level of 22\%
based on 126 data points). This result remains if we take all stars in
1b into account regardless of membership (then the correlation
coefficient is $-0.04$ at a significance level of 32\% based on 133
points). WH suggested that the trend they found could be due to the
superposition of members of subgroup 1a. It is not possible to
determine accurate spatial boundaries on the sky for the subgroups and
our data show that 1a is located at practically the same distance as
1b. Hence, we cannot exclude the possible contamination of the
membership list of 1b by 1a subgroup members.

\titleb {Ages of the subgroups}
Figure 3 shows the reddening-independent two-colour diagram $[B-U]$
vs.~$[B-L]$ for all the programme stars together with the line
representing the zero-age main-sequence (ZAMS), calculated from
theoretical stellar evolution models (see below). This diagram is the
observational counterpart of the \logt\ vs.~\logg\ diagram. Evolution
causes a star to move towards the lower left corner of the diagram, so
that the ZAMS is the upper right boundary of the data-points. Clearly,
the bulk of our programme stars lie at or near the ZAMS. GZL describe
the effects which cause a spread away from the ZAMS.

We have calculated isochrones based on the evolutionary models of
Schaller \ea\ (1992). We restricted ourselves to models with a
metallicity $Z=0.02$, and used only the core-hydrogen-burning and the
overall-contraction phases, which together cover the whole main
sequence strip in the HR-diagram. The isochrones were calculated by
means of the procedure described in section 4.2 of GZL.

The isochrones were used to determine the {\it nuclear age} of the
subgroups of Orion OB1, i.e., the time it took the stars to reach
their current evolutionary state measured from their arrival on the
ZAMS. The accuracy of the {\it absolute\/} age determined by isochrone
fitting depends on the calibration of the stellar models on which the
isochrones are based. The {\it relative\/} age of different subgroups
can often be determined more accurately, because the evolutionary
models are internally consistent.

\begfig 8cm
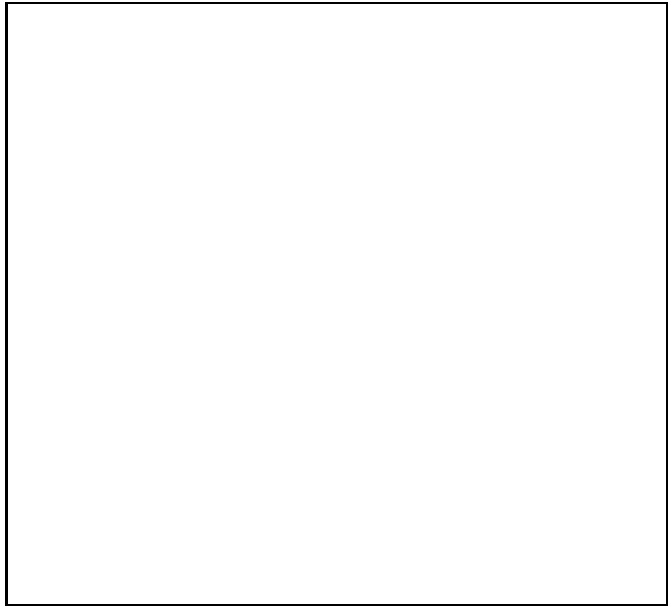
\figure {4a}{
Hertzsprung-Russell diagram of \logg\ vs.~\logt\ for subgroup 1a. The
solid line is the theoretical isochrone that represents the age of the
subgroup
}
\endfig

\begfig 8cm
\figure {4b}{
Same as 4a for subgroup 1b
}
\endfig

\begfig 8cm
\figure {4c}{
Same as 4a for subgroup 1c
}
\endfig

\begfig 8cm
\figure {4d}{
Same as 4a for subgroup 1d
}
\endfig

Isochrone fitting should preferably be done in the $[B-U]$ vs.~$[B-L]$
plane, because of the high accuracy of the photometry. At the
high-mass end, however, the theoretical isochrones do not overlap with
the Kurucz grid. Therefore, if the subgroup contains stars with masses
above $25$--$30\Msol$, we cannot calculate colours for isochrones
covering the turn-off from the main sequence. This occurs in subgroups
1b, 1c and 1d.  In this case, isochrone fitting is best done in the
\logg , \logt\ plane. The values of \logg\ and \logt\ for stars
outside the Kurucz grid have been taken from detailed spectroscopic
studies in the literature (section 3). Isochrone fitting in the \logl,
\logt\ plane would mean the introduction of extra errors caused by the
calibrations of absolute magnitude and bolometric correction.
Appendix A describes the adopted fitting procedure, and the Monte Carlo
simulations which were used to estimate the errors in the derived
ages. These simulations include the effects of the large errors in
\logg\ and the influence of the form of the initial mass
function. Chemically peculiar stars were excluded from the fitting
procedure. Figure 4 shows the \logg\ vs.~\logt\ diagrams for the four
subgroups of Orion OB1, and the isochrone corresponding to the age of
the subgroup.

Table 3 lists the derived ages of the subgroups, as well as the ages
given by WH, and by Blaauw (1991). WH used the intrinsic colours of
the bluest stars to determine the ages, by means of a technique
described by Maeder (1972). WH also give an extensive review of all
the age determinations done for Orion OB1 up to 1978. Many studies
show that ages decrease going from 1a to 1d, but we find that 1b is in
fact younger than 1c. This agrees with the determination by de Zeeuw
\& Brand (1985), based on $uvby\beta$ photometry. Further support
comes from the fact that subgroup 1b contains the most massive stars
in Orion OB1, which are $\zeta$ Ori A ($49\Msol$), $\delta$ Ori A
($45\Msol$) and $\epsilon$ Ori A ($42\Msol$). By contrast, the most
massive stars in subgroup 1c are $\iota$ Ori A ($36\Msol$) and
$\kappa$ Ori ($33\,
\Msol$) (Lamers \& Leitherer 1993; Vilkoviskij \& Tambovtseva 1992).
According to the Schaller \ea\ (1992) tracks, such masses put an upper
limit of $\sim\, 4\Myr$ on the age of subgroup 1b and a limit of
$\sim\, 5\Myr$ on the age of subgroup 1c. Note that also in the
association Sco OB2 the subgroups are not ordered along the sky
according to age. The oldest subgroup is located in between the two
younger ones (GZL, de Zeeuw \& Brand 1985).

\begtabfull
\tabcap {3} {Ages of the subgroups in Myr}
\halign{#\hfil&&\quad#\hfil\cr
\noalign{\hrule\medskip}
Subgroup&\vbluw&WH&Blaauw\cr
\omit&\omit&\omit&(1991)\cr
\noalign{\medskip\hrule\medskip}
	a&$11.4\pm 1.9$&7.9&12\cr
	b&$1.7\pm 1.1$&5.1&7\cr
	c&$4.6^{+1.8}_{-2.1}$&3.7&3\cr
	d&$<1.0$&$<0.5$&0\cr
\noalign{\medskip\hrule\medskip}}
\endtab

Age determinations based on kinematics were made by several
authors. This is usually done by tracing the projected paths of the
stars backwards in time from their current position, at a constant
velocity given by the observed proper motion. The time elapsed since
the moment that the member stars covered the minimum extent on the sky
is then taken as the {\it kinematic age}. In this way, Lesh (1968)
derived an age of $4.5\times 10^6\yr (\pm 30\%)$ for subgroup
1a. Blaauw (1961) found a kinematic age of 2.2--$4.9\times 10^6\yr$
for 1b, based on the motions of the three famous runaway stars, AE
Aur, $\mu$ Col and 53 Ari. We return to the question of the runaway
stars in section 4.5.  In section 5 we discuss the ages of the
subgroups in the context of sequential star formation.

\begfigwid 12cm
\figure {5}{
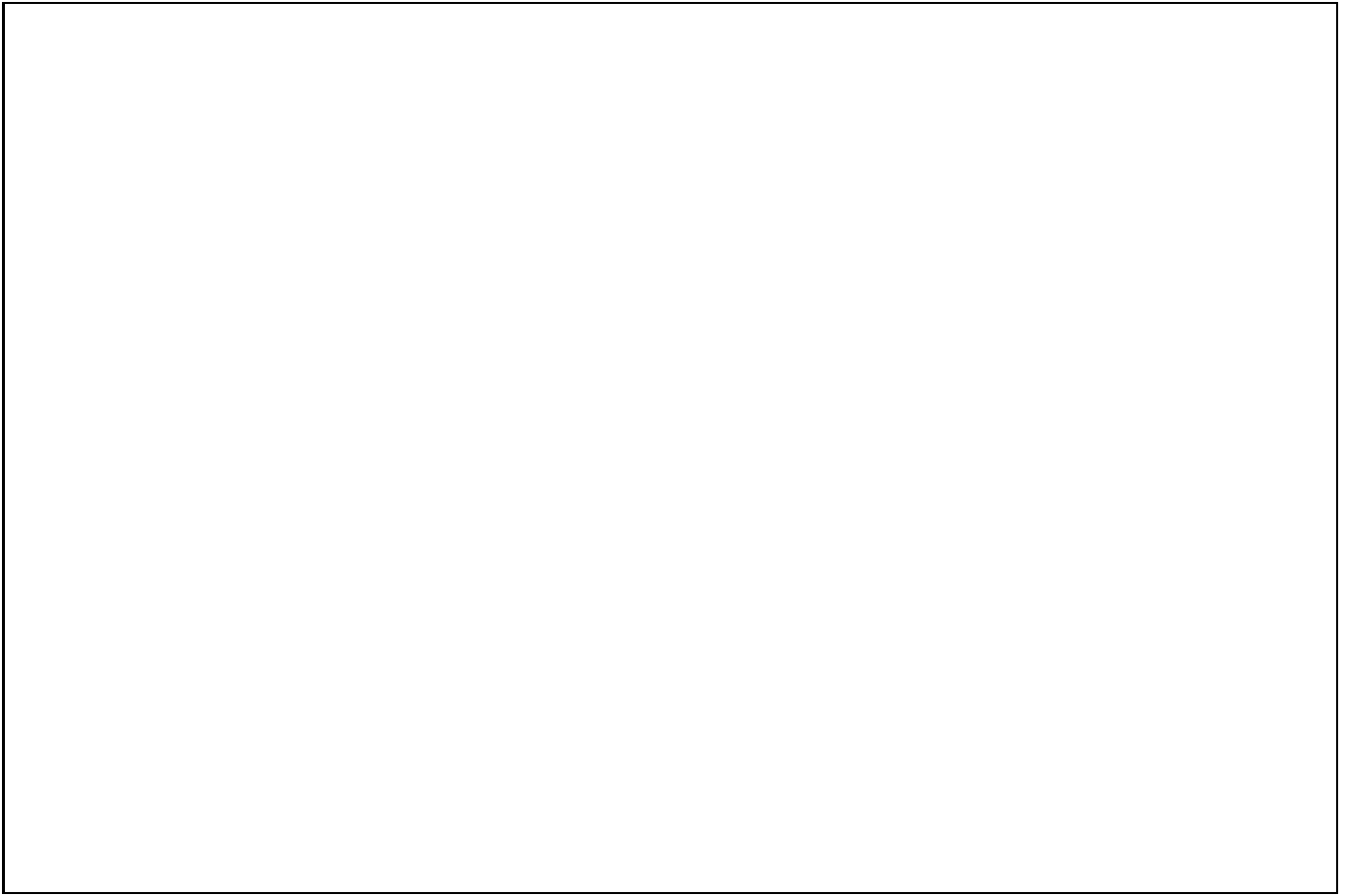
Grey scale map of the IRAS $100\um$ intensity in the Orion region. The
lowest intensities are indicated in black and the highest are white,
Galactic coordinates are used. The positions of the program stars are
indicated by open circles. Their size indicates the magnitude of the
visual extinction }
\endfig

\titleb {Visual extinctions and the gas/dust content of the Orion
region}
Figure 5 is a map of the IRAS $100\um$ emission in the Orion region.
This map is a mosaic of plates from the IRAS Sky Survey Atlas
(Wheelock \ea\ 1993). The superimposed circles denote the positions of
the programme stars. The size of a circle indicates the visual
extinction \av\ of the star, which is caused by foreground material.
Only non-peculiar stars and binaries for which we have photometry of
both components, or primaries unaffected by their companions, are
included in this figure. The Orion A and B clouds and a number of well
known dust/\HII regions are clearly outlined in this map. The Orion
nebula ($\ell\approx 209^\circ, b\approx -19.5^\circ$), the Horsehead
nebula, NGC 2023/NGC 2024 ($\ell\approx 206.5^\circ, b\approx
-16.5^\circ$), and NGC 2064 ($\ell\approx 205.5^\circ, b\approx
-14.5^\circ$) can all be seen. For the purpose of the discussion below
we shall define the region where the Orion A and B clouds are located
as follows. The region is enclosed between two lines. One line runs
from $(\ell ,b)=(\deg{217},\deg{-21})$ to $(\deg{203},\deg{-21})$ and
from this last point to $(\deg{202},\deg{-12})$. The other line runs
from $(\ell ,b)=(\deg{217},\deg{-17})$ to $(\deg{210},\deg{-17})$ and
from the last point to $(\deg{207},\deg{-12})$.  A small part of the
region that contains our programme stars lies within the ecliptic
latitude band ($\vert\lambda \vert <20^\circ$) where zodiacal emission
contaminates the IRAS fluxes. However, the contribution of zodiacal
emission is expected to be small in the $100\um$ band (e.g., Beichman
1987; Rowan-Robinson \ea\ 1991).

Inspection of Fig.~5 reveals that the correlation between \av\ and the
$100\um$ emission is not very tight. Many stars in the region of Orion
A and B have low \av . This can be due to: 1) the presence of small
scale density enhancements, so that different lines of sight probe
different dust columns, and 2) stars being at different depths in the
cloud. This second possibility suggests that we can combine the
knowledge of the distances and visual extinctions to the stars with
the $100\um$ emission, in order to constrain the distances to and the
depths of the Orion A and B clouds. Stars located in front of the
molecular clouds are only reddened by foreground extinction while
stars behind the clouds are also reddened by cloud material. Because
\av\ only measures the dust column in front of the star and the
$100\um$ emission measures the whole dust column along the line of
sight, one expects that the values for \av\ and $100\um$ are
essentially the same for stars behind the cloud. It follows that the
distance to the near edge of the clouds can be derived from an
analysis of
\av\ vs.~distance modulus (see GZL) and the far edge can be derived from the
difference between the observed $100\um$ emission and that predicted by \av\
(e.g., De Geus \& Burton 1991).

It is difficult to apply this procedure to Orion. The Orion A and B
clouds are very clumpy and filamentary, as is evident from studies in
optically thin lines (Bally \ea\ 1987). Stars located behind the
clouds may therefore have low extinction, due to the very small beam
of the photometric observations. The IRAS beam, on the other hand, is
much larger than the photometric beam, so the $100\um$ emission
contains contributions from a wider dust column around the line of
sight towards the star. However, if the sample of stars projected on
the clouds is large enough, one might be able to constrain the
distance and depth of the clouds in a statistical sense.

\begfig 8cm
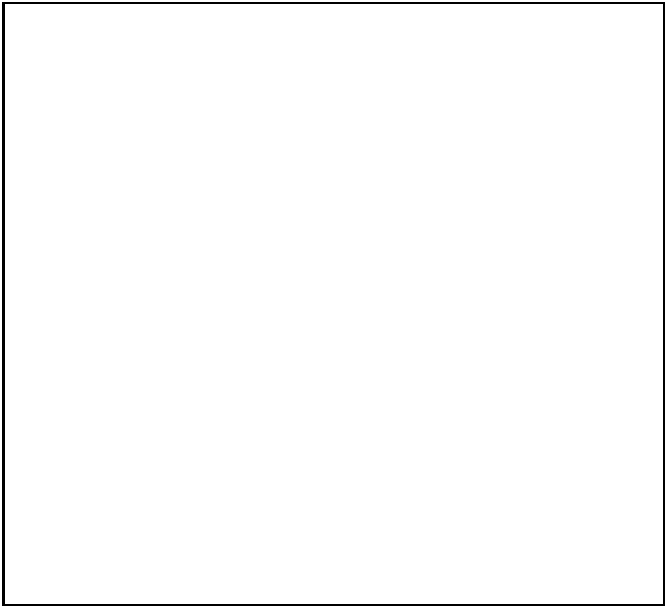
\figure {6a}{
Visual extinction vs.\ distance modulus for all the program stars
projected on the Orion A and B clouds. The line is the completeness
limit for B5{\sc v} stars
}
\endfig

\begfig 8cm
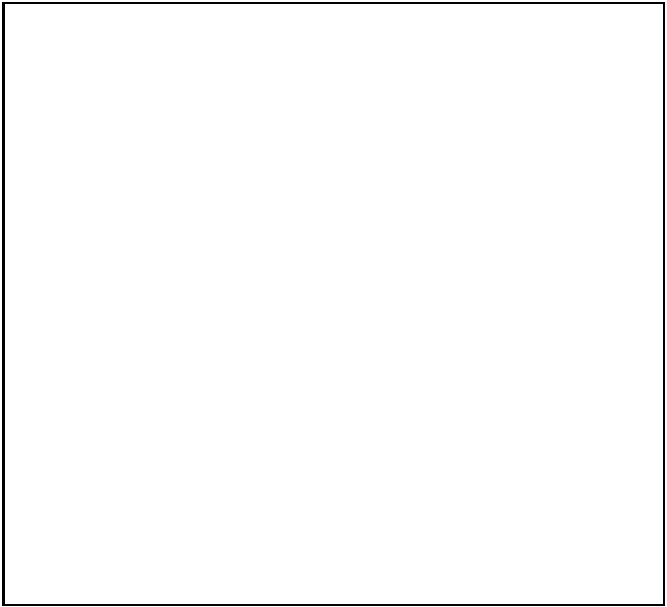
\figure {6b}{
Same as 6a for stars located away from the clouds
}
\endfig

\begfig 8cm
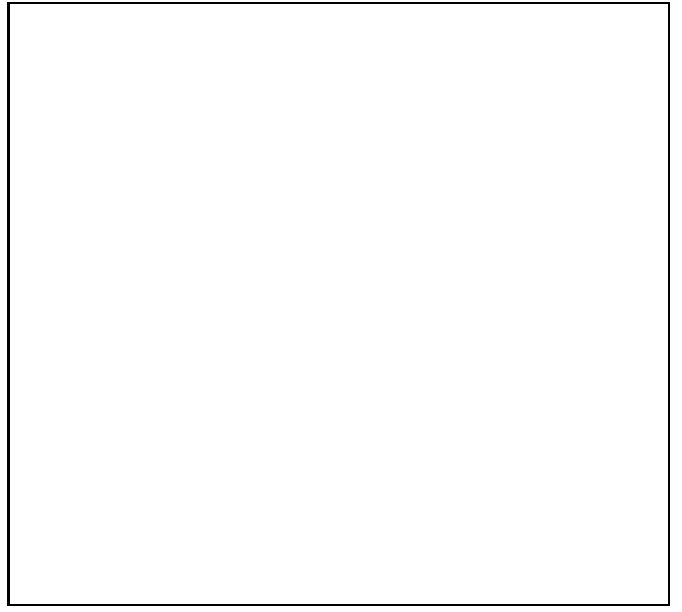
\figure {7a}{
\av\ vs.~$100\um$ intensity for the stars projected on the Orion A
and B clouds. The solid line is the left-best (eye) fit to the data
points as described in the text
}
\endfig

\begfig 8cm
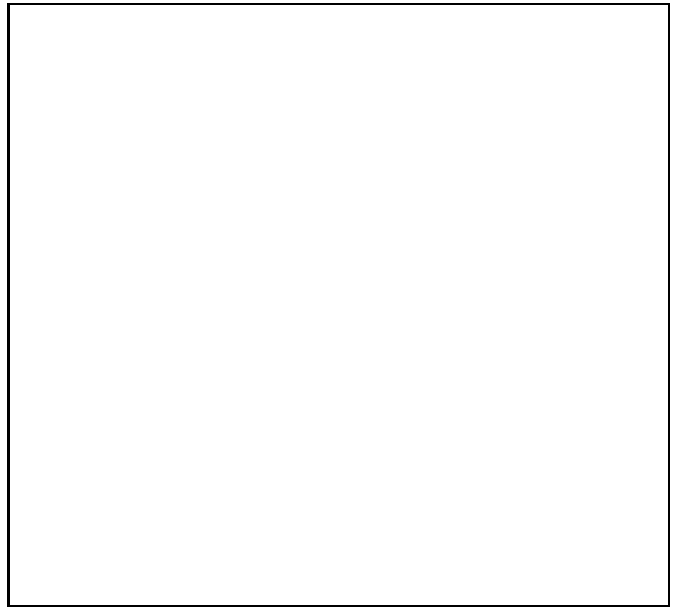
\figure {7b}{
The residual component of $I_{100}$ (corrected for the component
traced by \av ) vs.~the distance of the stars. The drop in the
residual component at $\sim 500\pc$ indicates the distance of the far
edge of the Orion A and B clouds (see text)
}
\endfig

We divided the programme stars in two groups: stars projected on Orion
A and B and stars located away from the clouds. Stars projected on the
clouds are located in the area that defines Orion A and B and stars
located away from the clouds are located at $b<\deg{-21}$ and at
$b<\deg{-16}$ for $\ell <\deg{203}$. Table 1 indicates which stars
were used in this analysis. Figures 6a and b show
\av\ vs.~distance modulus for the two groups. The influence of the
magnitude limit of the observations ($\sim 11\magn$) is evident in
both panels. The completeness limit for B5{\sc v} stars is drawn in
both plots. There is a difference between the two groups. Stars
located away from the clouds show a gradual fading out at large
distance moduli. For the stars projected on Orion A and B there is a
sharp limit at $\sim 9\magn$. The limit is at $\sim 8\moverdot 5$ for
the stars with low \av .  This suggests that we are looking into or
through the cloud. From Figure 6b it is clear that there is foreground
extinction at the distance of Orion OB1. Judging from the distribution
of \av\ values, the average foreground extinction is $\sim 0\moverdot
3$--$0\moverdot 4$. Therefore in order to estimate the distance of
the near edge of the clouds we have to look for the distance where
\av\ starts to rise significantly above $0\moverdot 4$. Judging from
Fig.~6a, this occurs roughly at a distance modulus of $7.5\pm
0\moverdot 5$ ($\approx 320\pc$).

Now we turn to the far edge of the clouds. Figure 7a shows \av\
vs.~$100\um$ intensity for the stars projected on Orion A and B. The
line is the left-best (eye) fit to the data and is an indication of
the general relation between \av\ and the $100\um$ emission. The line
is given by:
$$I_{100}=8+30A_V ({\rm MJy\, sr^{-1}})\; .                   \eqno (5)$$
This relation tells us which part of the dust column along the line of
sight is measured by \av\ (De Geus \& Burton 1991). Subtracting the
$100\um$ emission given by this relation from the observed emission
indicates which stars lie behind the clouds. As explained above, for
these stars the dust column measured by \av\ should be the same as
that measured by the $100\um$ emission. The results are shown in
Fig.~7b, where the ``corrected'' $100\um$ emission is plotted
vs.~distance. This figure shows that the far edge of the clouds is at
roughly $500\pm 30\pc$. However, this is based on a small number of
stars at large distances.

In conclusion, the distances to the near and far edge of the clouds
are $\sim 320$ and $\sim 500\pc$, placing the clouds at a distance of
$410\pc$ with an overall depth of approximately $180\pc$. The Orion
clouds span an angle of $\sim 16^\circ$ on the sky, which implies a
tangential extent of roughly $120\pc$. The tangential extent is
consistent, within the error bars, with the depth of the clouds. An
improved estimate of the depth and distance of the clouds and a
possible detection of differences in distance to the various parts of
the clouds requires a larger sample of stars in the direction of the
clouds and the sample should include fainter stars.

\titleb {Initial mass function}
The mass of a star is related to \logl, \logt\ and \logg\ by:
$$ \log {M\over {\Msol}}=\lg +\log {L\over {{\rm L}_\odot}}-4\lt +10.62\;
, \eqno(6)$$
where the constant in the equation is a combination of the effective
temperature and surface gravity (in ${\rm cm}\, {\rm s}^{-2}$) of the
Sun. The derived distribution of masses in a subgroup gives the
present day mass function (PDMF). The limits of the mass interval on
which the PDMF is complete are set by the maximum mass consistent with
the age of the subgroup, and by the minimum mass at which the
observations are complete. The initial mass function (IMF) of a
subgroup was determined by fitting the PDMF on the observed mass
interval to a number of functional forms $\xi(\log M)\diff\log M$ with
the aid of a Kolmogorov-Smirnoff (KS) test. The underlying assumption
is that the star formation process or, conversely, the fragmentation
of a molecular cloud, can be described as a stochastic process (Cayrel
1990).

In order to investigate the merits of determining the IMF with the
KS-test we performed extensive Monte Carlo simulations, described in
Appendix B. We find that a single power law of the form $\xi (\log
M)=AM^{-B}$ with $B=1.7\pm 0.2$ gives an adequate description of the
IMF of each of the subgroups. This IMF is consistent with the results
obtained by Claudius \& Grosb\o l (1980), who find $B=1.9\pm 0.2$.
Table 4 gives for each subgroup the mass interval on which the PDMF
was constructed, the number of stars on this interval and the number
of stars between 2 and $120\Msol$, according to the best fit
IMF. Note, that in contrast to Blaauw (1991), we find that subgroup 1c
is the richest subgroup (in number of stars) and that all the
subgroups in Orion OB1 are comparable in richness to the
Upper-Scorpius and Upper-Centaurus-Lupus subgroups of Sco OB2 (De Geus
1992).

\begtabfull
\tabcap {4} {Total number of stars between 2 and $120\Msol$. The mass
range of the PDMF and the number of stars therein is indicated}
\halign{#\hfil&&\quad#\hfil\cr
\noalign{\hrule\medskip}
Subgroup&mass range of PDMF&${\rm N}_{\rm\scriptscriptstyle PDMF}$&${\rm
N}_{\rm tot}$\cr
\omit&($\Msol$)&\omit&\omit\cr
\noalign{\medskip\hrule\medskip}
	a&4--15&53&190\cr
	b&4--120&45&150\cr
	c&7--36&23&210\cr
\noalign{\medskip\hrule\medskip}}
\endtab

Undetected duplicity or spectral peculiarities introduce errors in the
masses derived by means of Eq.\ (6). Not taking into account the
components of multiple systems can lead to a severe underestimate of
the number of observed stars contributing to the IMF in the adopted
mass interval. For this reason we took masses for the known multiple
stars from the literature, or we estimated them based on the magnitude
difference with respect to the primary. These mass estimates were used
to determine which components of a multiple system actually contribute
to the PDMF in the observed mass interval. We find that the number of
known multiple systems in the Orion OB1 subgroups amounts to $\approx
30\%$ of the total number of stars in our sample.

We estimated the contribution of undetected spectroscopic binaries
(SB) to the PDMF by assuming that 30\% of all stellar systems are SBs
(e.g., Blaauw 1991).  We calculated a statistical correction by
assuming a distribution of mass ratios as given by Hoogeveen (1992)
for SB systems with a primary of spectral type B. The resulting
correction for the presence of SB systems is rather small, amounting
to typically 2 objects per subgroup, and can be ignored in the process
of determining the IMF. The number of undetected SB systems is
expected to be small because of the limited mass range on which the
PDMF is determined. This means that no SB systems with secondaries of
mass lower than $4\Msol$ are considered and also SBs with primaries
with a mass below this limit are excluded. Peculiar stars were checked
individually and if their derived masses were obviously in error
compared to their spectral type, an estimate of the mass was made
based on the spectral type of these stars. For subgroup 1d we did not
attempt to derive an IMF because the photometry is not very reliable
in this area and the observations are incomplete.

\titleb {Energy output}
The total energy output of an association over its lifetime is
dominated by the OB stars (De Geus, 1992).  To estimate the energy
output of these stars, we have to distinguish the stars still present
in the association from those that have already disappeared from
it. For the former only the energy output through stellar winds during
their main-sequence phase up to the age of the association is
important. For the latter it is assumed that they have already gone
through their entire evolution and have expired (or ``evaporated'') as
supernovae. These stars make important contributions to the energy
output by the various stellar wind phases during their lifetime and by
the final supernova explosion.

Massive stars experience a number of mass-loss phases during their
evolution: the main-sequence or OB phase, the luminous blue variable
phase, the red supergiant phase, and the Wolf-Rayet (WR) phase
(Leitherer \ea\ 1992). The occurrence of the last three evolutionary
stages, and hence their importance for the energy input into the ISM,
depends on the initial mass. Leitherer \ea\ (1992) studied the
deposition of mass, momentum and energy into the ISM by massive stars
in order to apply the results to starburst galaxies. In their Fig.~3
the contributions of the four evolutionary stages to the energy
deposition are shown for a model assuming an instantaneous starburst,
which is appropriate for an OB association. We conclude that only the
OB and the WR phases have to be considered in order to calculate the
stellar wind energy output of the association.

The contributions of the OB phase were calculated by deriving the
mechanical luminosity in the wind, $\dot E_{\rm wind}$, from stellar
parameters as a function of time for each star, and integrating this
quantity over the main-sequence lifetime of the stars already
evaporated or over the age of the association for stars still
present. Due to the fact that there are still O stars left in the
subgroups of Orion OB1, the contributions of the stars still present
are significant, in contrast to the case of, e.g., Sco OB2 (De Geus
1992).

\begfig 8cm
\figure {8}{
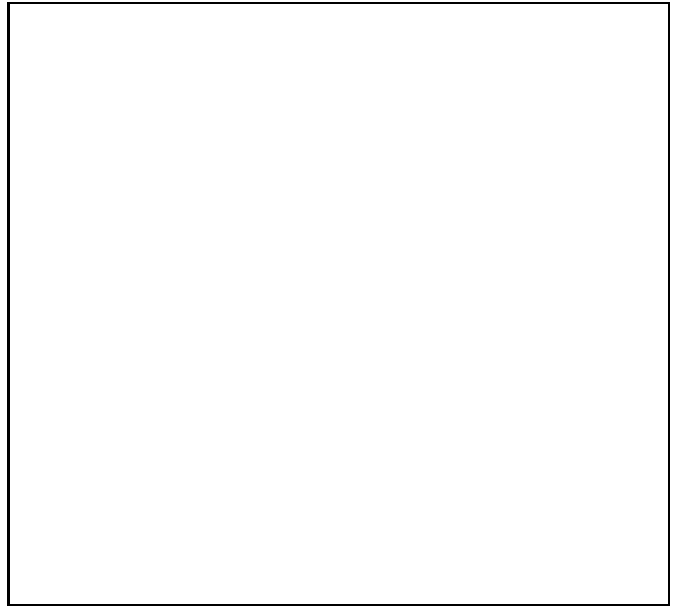
The number of stars that have exploded as supernovae as a function of
time for 1000 realizations of a synthetic association with a single
power law IMF; $\xi (\log M)=AM^{-1.7}$, containing 210 stars between 2
and $120\Msol$. The solid line is the average number of supernovae,
the dotted lines are the rms spreads around that average and the dashed
lines represent the maximum and minimum number of supernovae that
occurred
}
\endfig

To calculate $\dot E_{\rm wind}$, the mass-loss rate, \mdot , and the
terminal velocity of the wind, \vterm , are required. For very
luminous O stars with $\loglum > 5.0$ and $\lt > 4.45$, we used the
results from a recent study by Lamers \& Leitherer (1993) to calculate
the mass-loss rate as a function of \logl\ and \logt. The terminal
velocity, \vterm , was calculated from the escape velocity of these
stars with the aid of the simple relation, ${v_\infty}/{v_{\rm
esc}}=2.8$ (Groenewegen \ea\ 1989). We used the ``recipe'' of
Kudritzki \ea\ (1989) to derive \mdot\ and \vterm\ for late O and
early B stars. Lamers \& Leitherer find that this recipe predicts
mass-loss rates that are on average a factor of two lower than those
observed, with the precise ratio a function of the observed mean
density in the wind. The predicted terminal velocities are on average
a factor $1.4$ larger than observed (see also Groenewegen \ea\ 1989).
We therefore corrected the mass loss rates calculated with the
Kudritzki \ea\ algorithm by means of Eq.\ (20) of Lamers \& Leitherer
(1993), and we multiplied the predicted terminal velocities by 0.7.

We assumed that stars with masses above $25 \Msol$ go through a WR
phase. We adopted an average mass-loss of $2\times 10^{-5} \Msol /{\rm
yr}$ in the WR phase, and a terminal velocity of $\approx 2000\kms$.
The duration of the WR phase varies between $2.5\times 10^5\yr$ and
$6\times 10^5\yr$ depending on the initial mass of the star and the
mass-loss rate (Schaller \ea\ 1992). For an average duration of
$4\times 10^5\yr$, each WR star contributes $3\pm 2\times 10^{50} {\rm
ergs}$.

Finally, we assumed that each SN explosion produces $10^{51}\, {\rm
ergs}$. Roughly 20\% of the total energy output of an association will
contribute to the kinetic energy of \HI shells surrounding it (e.g.,
Weaver \ea\ 1977).

\begfig 8cm
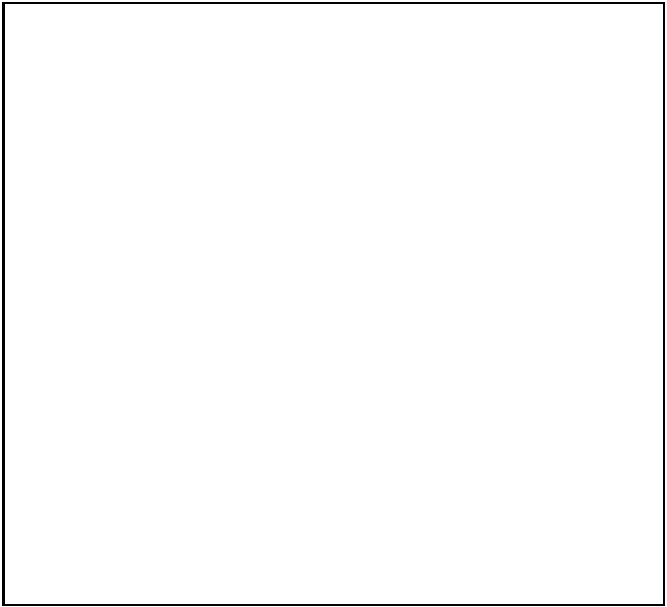
\figure {9a}{
Distribution of wind-energy outputs obtained in 1000 realizations of a
synthetic association. The synthetic association contains the same
number of stars between 2 and $120\Msol$ as subgroup 1a (Table~4) and
has the same IMF. Note the long tail towards large energy outputs }
\endfig

\begfig 8cm
\figure {9b}{
Same as 9a for the parameters of subgroup 1b
}
\endfig

\begfig 8cm
\figure {9c}{
Same as 9a for the parameters of subgroup 1c
}
\endfig

Comparison of the different contributions shows that supernovae and
massive O-stars dominate the energy output of the subgroups. As a
result, the total energy released by the association is very sensitive
to the number of massive stars that were formed or, conversely, to the
upper mass cutoff in the IMF.  If the process of forming an OB
association with a certain IMF is viewed simply as generating a
population of stars from a probability distribution given by the IMF,
then the number of high mass stars will be very ``noisy''.  This is
due to the relatively small number of stars formed in an association
and the slope of the IMF.

The effects of this noise at the high mass end of the IMF were
investigated with the aid of Monte Carlo simulations. We generated
synthetic associations according to the power law IMF with index
$-1.7$ that we derived for the Orion OB1 subgroups (section 4.4). We
assumed the IMF to hold for the range 2--$120\Msol$. The real upper
mass cutoff is not known, so allowing stars up to $120 \Msol$ to form
will indicate the full range of possibilities.

Figure 8 shows the cumulative number of evaporated stars as a function
of the age of the association, assuming 210 stars between 2 and $120\,
\Msol$ were formed. This figure shows that the number of supernovae
which occur during the early lifetime of an association has a large
statistical uncertainty. The relative uncertainty is largest during
the first $10\Myr$; which is precisely the range relevant to this
work.

The wind-energy output was calculated by combining the simulations
with the calculation of wind-energies for individual stars as
described above. The results are shown in Fig.~9, which gives the
distribution of energy outputs for each of the three subgroups. All
three distributions have a definite peak but they are very asymmetric
with large tails out to high energies. These tails are caused by the
fact that the total energy output from stellar winds is a very strong
function of the mass of the most massive stars that were actually
formed in the subgroup. This is shown for one subgroup (1a) in
Fig.~10, where the energy output is plotted vs.~the mass of the most
massive star in the subgroup. Even the existence of one very massive
star can have a large effect on the total energy output. It follows
that a simple average of the distribution of wind energies is not a
good estimator of the expectation value of the total energy output. It
is better in this case to take the median, which is less sensitive to
tails in the distribution. The error bars will also be asymmetric.

\begfig 8cm
\figure {10}{
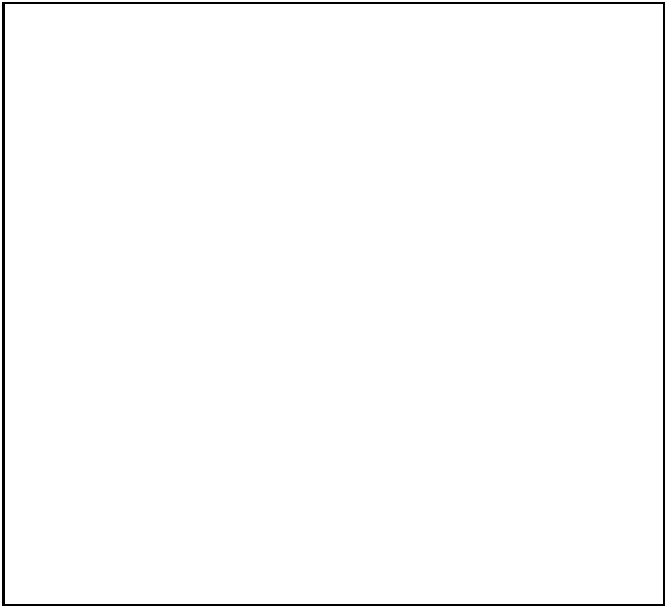
Wind-energy output vs.\ mass of the most massive star formed for 1000
realizations of a synthetic association with parameters appropriate
for subgroup 1a of Orion OB1
}
\endfig

The energy outputs that we derive for the subgroups in Orion OB1 are
shown in Table 5. In these calculations we did not take into account
the fact that some stars might still be embedded in their own \HII
region and therefore do not contribute to the energy output of the
association as a whole. This is, for instance, clearly the case for
the stars in subgroup 1d. To estimate an upper limit on the magnitude
of this error source we assume that a star is embedded in its own
\HII region throughout its main-sequence evolution. This means that we
cannot count the contribution of the stellar wind of such a star in
the total energy output. The most massive stars in Orion range in mass
from 30--$50\Msol$. Their individual energy outputs on the main
sequence range from 1--$3\times 10^{50} {\rm ergs}$.

\begtabfull
\tabcap{5} {Energy outputs in $10^{50}$ ergs}
\halign{#\hfil&&\quad#\hfil\cr
\noalign{\hrule\medskip}
Subgroup&SN&WR&OB&total\cr
\noalign{\medskip\hrule\medskip}
	a&$59\pm 25$&$7\pm 5$&$7^{+11}_{-3}$&$73^{+40}_{-33}$\cr
\noalign{\smallskip}
	b&$0 (<20)$&$0 (<6)$&$2^{+8}_{-1}$&$2^{+8}_{-1} (<28)$\cr
\noalign{\smallskip}
	c&$14\pm 12$&$4\pm 4$&$6^{+11}_{-4}$&$25^{+27}_{-19}$\cr
\noalign{\medskip\hrule\medskip}}
\endtab

We find that with our smaller inferred age for subgroup 1b, at most
two supernova events are likely to have occurred in it. There are
three known runaway stars that could have originated in 1b: $\mu$ Col,
AE Aur and 53 Ari. If one accepts the scenario that runaways are
produced by supernova explosions in binaries (Blaauw 1961), this would
mean that at least three supernova events have occurred in 1b. But for
the first two runaways it is not well established from which subgroup
they originated and this pair could be a candidate for the dynamical
ejection scenario (Gies \& Bolton 1986, Leonard \& Duncan 1988, 1990),
because of their opposite motions, nearly equal masses, speeds and
kinematic ages. The runaway 53 Ari could also have originated from 1a,
in agreement with its kinematic age of 7.3 million years (see Blaauw
1991, 1993 for more details on these runaways).

\titlea {The interaction of stars and the interstellar medium in Orion}
The large scale structures in the interstellar medium around Orion OB1
have been studied extensively by means of a variety of observational
techniques.  Much of the early observational material was summarized
by Goudis (1982). The large scale features include the Orion Molecular
Cloud complex (OMC) (Maddalena \ea\ 1986; Kutner \ea\ 1977), H$\alpha$
emission (including Barnard's loop) which extends all the way to the
Eridanus region 50$^\circ$ below the Galactic plane (Sivan 1974;
Reynolds \& Ogden 1979, hereafter RO) and a hole in the neutral
hydrogen distribution surrounded by \HI arcs showing expanding motions
(Heiles 1976; RO). The diffuse Galactic H$\alpha$ background intensity
is enhanced in this area, and the same is true also for the X-ray
intensity (e.g., RO; Burrows \ea\ 1993). Part of the expanding \HI
shell around the \HI cavity was detected also by Green (1991) and
Green \& Padman (1993). Their \HI observations extend to $\sim\,
35^\circ$ below the Galactic plane and include the Orion OB star
complex. Cowie
\ea\ (1979) measured ultraviolet absorption lines of various
ionization stages of C, N, Si and S for stars in or near Orion OB1,
and detected a high-velocity, low-ionization shell surrounding all
of the Orion area (Orion's Cloak), as well as a dense, clumpy
low-ionization shell expanding at smaller velocity. Part of this
dense shell may be the negative velocity clump (in \HI) described by
Green \& Padman (1993).

RO interpreted the various observed features as being caused by a
cavity of hot ionized gas surrounded by expanding shells of \HI. The
ionization of the cavity is maintained by the UV radiation from the
early-type stars in Orion OB1. They explained the existence of the
shell as due to a series of supernova explosions about $2\times 10^6$
years ago. Cowie \ea\ interpreted the high velocity shell as the
radiative shock of a supernova which occurred $3\times 10^5$ years
ago, and the high-column-density structures as the remnants of older
supernova events (2--$4\times 10^6$ years ago). Burrows \ea\ (1993)
present a model in which the observed X-ray enhancement is due to hot
gas which fills a cavity created by the stellar winds from Orion OB1.

We have shown in section 4 that the energy output from winds
contributes significantly to the total energy output of the Orion OB1
association. Bally \ea\ (1991) presented observational evidence for
ongoing interaction between the stars in Orion OB1 and its
surroundings. The overall morphology of the molecular gas around Orion
OB1 has a wind swept appearance (see Fig.~2 in Bally \ea ), which is
confirmed by several systematic patterns appearing in the morphology
and kinematics of the molecular gas in Orion. Large scale velocity
gradients point toward Orion OB1 and the densest parts of the
molecular clouds face the association. Furthermore, the most active
star formation is taking place in the dense gas closest to the OB
association. In this section we investigate whether the ionized cavity
containing Orion OB1 can be understood as part of an \HI bubble
created and maintained by the combined effect of the winds and
supernovae from the association. This is followed by a discussion on
sequential star formation in Orion OB1.

\titleb {A simple calculation}
As a first approximation for the calculation of the size and expansion
velocity of the shell surrounding Orion OB1 we use the standard theory of
wind-driven bubbles as given by Castor \ea\ (1975) and Weaver
\ea\ (1977). McCray \& Kafatos (1987) have shown that this theory can also be
used for OB associations in which SN occur. We first assume spherical
geometry and a uniform density ISM surrounding the \HI shell. Then the
shell radius $R_{\rm S}$ and expansion velocity $V_{\rm S}$ are given
by (Shull 1993):
$$R_{\rm S} =(26.2\pc)L_{36}^{1/5}n_0^{-1/5}t_6^{3/5}\; ,
                                                                 \eqno (7)$$
$$V_{\rm S} =(15.4\kms)L_{36}^{1/5}n_0^{-1/5}t_6^{-2/5}\; ,
                                                                 \eqno (8)$$
where $L_{36}$ is the mechanical luminosity of the association
averaged over its lifetime, expressed in units of $10^{36}\ergps$, and
$t_6$ is the age of the association in $10^6\yr$. The ambient number
density of the ISM is given by $n_0$, and has units of cm$^{-3}$.

The observed radius of the \HI cavity is $\sim140\pc$ and the
expansion velocity is $15$--$20\kms$ (RO). If we take $L_{36}$ and
$t_6$ appropriate for Orion OB1a (20 and $11.4$ respectively), and use
either $n_0=1\cc$ or $n_0=0.1\cc$, then the expected size of the
cavity is 210 or $330\pc$, while the expansion velocity is 11 or
$17\kms$.  So the radius of the
\HI bubble can be readily explained with the energy output of Orion
OB1a alone, but the expansion velocities are too low.  This estimate
assumes, however, that the \HI shells started expanding at the moment
the stars in 1a were formed. This is not a valid approximation, as the
stars have to clear their immediate surroundings (the parental
molecular cloud) before their winds can influence the surrounding
ISM. Not surprisingly, the dynamical age of the shells, as derived
from the observed expansion velocities and size of the bubble, is
$3$--$5\Myr$. If we take this as the value for $t_6$, and neglect the
fact that part of the calculated energy output has gone into the
disruption of the parental molecular cloud, then we find: $R_{\rm
S}=120$--$150\pc$, $V_{\rm S}=17$--$24\kms$ ($n_0=1\cc$) and $R_{\rm
S}=190$--$240\pc$, $V_{\rm S}=28$--$38\kms$ ($n_0=0.1\cc$), where
$L_{36}$ is again based on the output from subgroup 1a alone. The
energy contributions from 1b and 1c will compensate for the fact that
not all of the energy of 1a goes into the
\HI shells.

We conclude that in the spherical symmetry, constant ambient density
approximation, the observed size and expansion velocities are
accounted for by the stellar winds and supernovae from Orion OB1.

Note that in the model for the Orion-Eridanus bubble presented by
Burrows \ea\ (1993) the required stellar wind luminosities are
3--$10\times 10^{36}\ergps$, depending on the distance of the X-ray
emitting gas. The stellar wind luminosities we find for the subgroups
of the association (4--$7\times 10^{36}\ergps$) are all in this
range. Thus the total stellar wind luminosity from Orion OB1 is enough
to account for the observed X-ray enhancement.

\titleb {The effects of a non-uniform ISM; blow out?}
The assumptions of spherical symmetry and a constant ambient density
clearly do not hold for Orion OB1. The star complex is located at
$\sim 130 \pc$ below the Galactic plane, and we therefore have to take
into account the density gradient in the \HI distribution
perpendicular to the plane of the Milky Way.  Furthermore, the \HI
shells extend to $\sim 300 \pc$ below the Galactic plane, and the
association is not in the center of the shells (RO; Bally \ea\ 1991).

No simple analytical results have been derived for the case of \HI
bubbles expanding in stratified media, but various numerical
simulations have been performed (Mac Low \& McCray 1988; Tomisaka \&
Ikeuchi 1986; Tenorio-Tagle \& Bodenheimer 1988). These simulations
show that the resulting bubbles are not spherically symmetric, but
tend to be larger in the direction perpendicular to the Galactic
plane. If the energy input is large enough, a ``blow out'' phenomenon
might occur: in this case the superbubble grows in size until it
breaks out of the Galactic \HI layer, thereby discharging its interior
energy into the Galactic halo.  For energy sources not located in the
plane of the Galaxy, one-sided blow out can occur.

How does the Orion-Eridanus bubble fit into this picture? The data
presented in Fig.~1 of RO and the schematic picture presented by Bally
\ea\ (1991) show clearly that the appearance of the Orion-Eridanus
bubble is in good qualitative agreement with the results of numerical
simulations in which the energy source of the bubble is placed below
the Galactic plane (see Figs.~5 and 8 in Mac Low \& McCray 1988).

Is it possible that blow out will occur in the future? Mac Low \&
McCray (1988) discuss the bubble dynamics in two stratified media: an
exponential distribution with a scale height of $100\pc$, and a more
realistic hybrid model consisting of a Gaussian cloud layer with a
scale height of $135\pc$ and an exponential \HI layer with a scale
height of $500\pc$ (Lockman \ea\ 1986). The dynamical parameter that
determines whether a bubble will blow out or not is the dimensionless
quantity $D$ defined as:
$$D\approx 940 L_{38} \biggl({H\over{100\pc}}\biggr)^{-2}
                      \biggl({P_{\rm e}\over {10^4\, k\, {\rm dyne}\,
                      {\rm cm}^{-2}}}\biggr)^{-3/2}n_0^{1/2}\; , \eqno(9)$$
where $L_{38}$ is the mechanical luminosity of the association in
$10^{38}\ergps$, $H$ is the scale height, and $P_{\rm e}$ is the
pressure of the external ISM. The value of $D$ appropriate for Orion
OB1 ranges from $\sim\, 60$ to $\sim\, 700$, depending on the
calculation of $L_{38}$ and the assumed ambient density. Figure 6 of
Mac Low \& McCray shows which combinations of $D$ and the bubble
dynamical time scale result in blow-out for the case of an exponential
medium. The dynamical time scale of the Orion-Eridanus bubble is
3--$5\Myr$, which would demand values of $D\approx 100$ for blow
out. So, in this case, blow out of the Orion-Eridanus bubble may
occur. The location at about 1.3 exponential scale heights from the
Galactic plane also favors one-sided blow out. However, the effective
scale height of the ISM is larger in the more realistic hybrid model,
and as a result much higher values of $D$ ($\sim\, 1000$) are required
for blow out (see also Fig.~8 in Mac Low \& McCray). This suggests
that the Orion-Eridanus bubble will not blow out of the Galactic \HI
layer.

\titleb {Sequential star formation in Orion} Star formation is
currently taking place in the OMC, at sites that are adjacent to the
subgroups of the association. The ambient molecular gas shows signs of
interaction with the Orion OB1 stars. This is indicative of sequential
star formation, for which there are a number of morphological
signatures. These signatures were already noted in various
associations, including Orion OB1, by Blaauw (1964). They have
recently been listed by Elmegreen (1992). They are: (i) regions of
star formation with ages separated by several $\Myr$ and with
distances between them of $\sim 10$--$50\pc$, (ii) near the youngest
subgroup there is still a substantial amount of gas and very little or
no gas is left near the oldest subgroup, (iii) a velocity difference
between the older and younger subgroups of $\sim 5$--$10\kms$,
observed in either the velocities of the stars or gas or both.

In Orion OB1 all three signatures can be found. The youngest subgroup,
1d, is still closely associated with the gas (being embedded in its
own \HII region), while the oldest, 1a, is not associated with any
gas. The distances between the four subgroups are of the order of
30--50$\pc$ and the age differences are 3--10$\Myr$. No differences in
radial velocity have been found so far between the subgroups (WH;
Morrell \& Levato 1991). The average radial velocity for the whole
association is $23\kms$ (Morrell \& Levato 1992), whereas the bulk
velocity of the gas in the Orion A and B molecular clouds is
3--$12\kms$ (Maddalena \ea\ 1986). A possible explanation is that the
stars were formed while the gas was moving at higher velocities. The
stars retained this higher velocity while the gas slowed down. The
difference in present stellar and gas velocities is also related to
the question of the origin of the first subgroup to which we return
below.

Judging from the ages for the individual subgroups there is no linear
progression of star formation from 1a to 1d. Subgroup 1c was formed
before 1b and yet 1b is closer to 1a in projection, although the
difference is rather small ($\sim 10$--$20\pc$).  This can possibly be
explained by the geometry of the system: subgroup 1c formed in a part
of the OMC that was closer to 1a in the past and has since moved to
its present position. At a speed of 3--$6\kms$ it could have moved the
required 10--$20\pc$ in $\sim 3\Myr$.  Another aspect is that 1c seems
more closely associated to the gas than does 1b, but this could again
be a projection effect, with 1c located in front of the Orion B cloud
and the \HII regions in it. It is very likely that subgroups 1b and 1c
are now themselves triggering star formation in various sites in the
Orion A and B clouds.

Where did the oldest subgroup, 1a, come from?  Several explanations
have been proposed, which also involve the formation of the OMC. One
possibility is that the OMC was formed as the result of the impact of
a high velocity cloud (HVC) on the Galactic disk (Franco \ea\ 1988).
The HVC approached the disk from below the Galactic plane and as a
consequence of the collision a self-gravitating remnant was left which
oscillated once through the Galactic plane and now constitutes the
OMC.  The fact that the impact took place in the south then explains
the position of subgroup 1a.

Another explanation for the existence of the OMC is that it is part of
the gas swept up by the expanding ring of gas known as Lindblad's ring
(Lindblad \ea\ 1973; Olano 1982). The gas swept up in this ring was
subjected to Galactic tidal forces and fragmented into a number of
clouds, among which the OMC, in which star formation occurred
(Elmegreen 1992, Fig.~15). The initially higher expansion velocity of
the ring (Olano 1982) might be the reason why the stars in Orion OB1
move at higher radial velocities than the gas in the OMC. It is
interesting to note in this context that the ages of the oldest
subgroups in two other associations linked with the ring (Sco OB2 and
Per OB2) are similar to the age of Orion OB1a. The existence of the
ring itself might be explained as a consequence of the deposition of
energy into the ISM by the old Cas-Tau association (Elmegreen 1992;
Blaauw 1991). Alternatively, the formation of the ring might be
related to the origin of Gould's belt. Lindblad's ring is associated
with Gould's belt (e.g., Olano 1982) and it has recently been
suggested that the belt was formed as the consequence of the impact of
a high velocity cloud on the galactic disk (Comer\' on \& Torra
1992). Such an impact could also lead to the formation of Lindblad's
ring (e.g., Elmegreen 1992).

We conclude that past geometries of the system of stars and gas can
explain our finding that subgroup 1b is younger than 1c. The formation
of 1a might be the consequence of the impact of a high velocity cloud
on the Galactic disk or it might have formed when the expanding
Lindblad ring fragmented into molecular clouds.

\titlea {Conclusions and  future work} We derived  physical parameters
for candidate members of the Orion OB1 association, based on \vbluw
photometry.  Distances to the stars and the subgroups of the
association were calculated and it was shown that in subgroup 1b there
is no increase in distance from west to east. New photometric members
of Orion OB1 were identified based on the distances to the subgroups
and the rms spreads in them, calculated from the distances of known
members.

\begfigwid 12cm
\figure {11}{
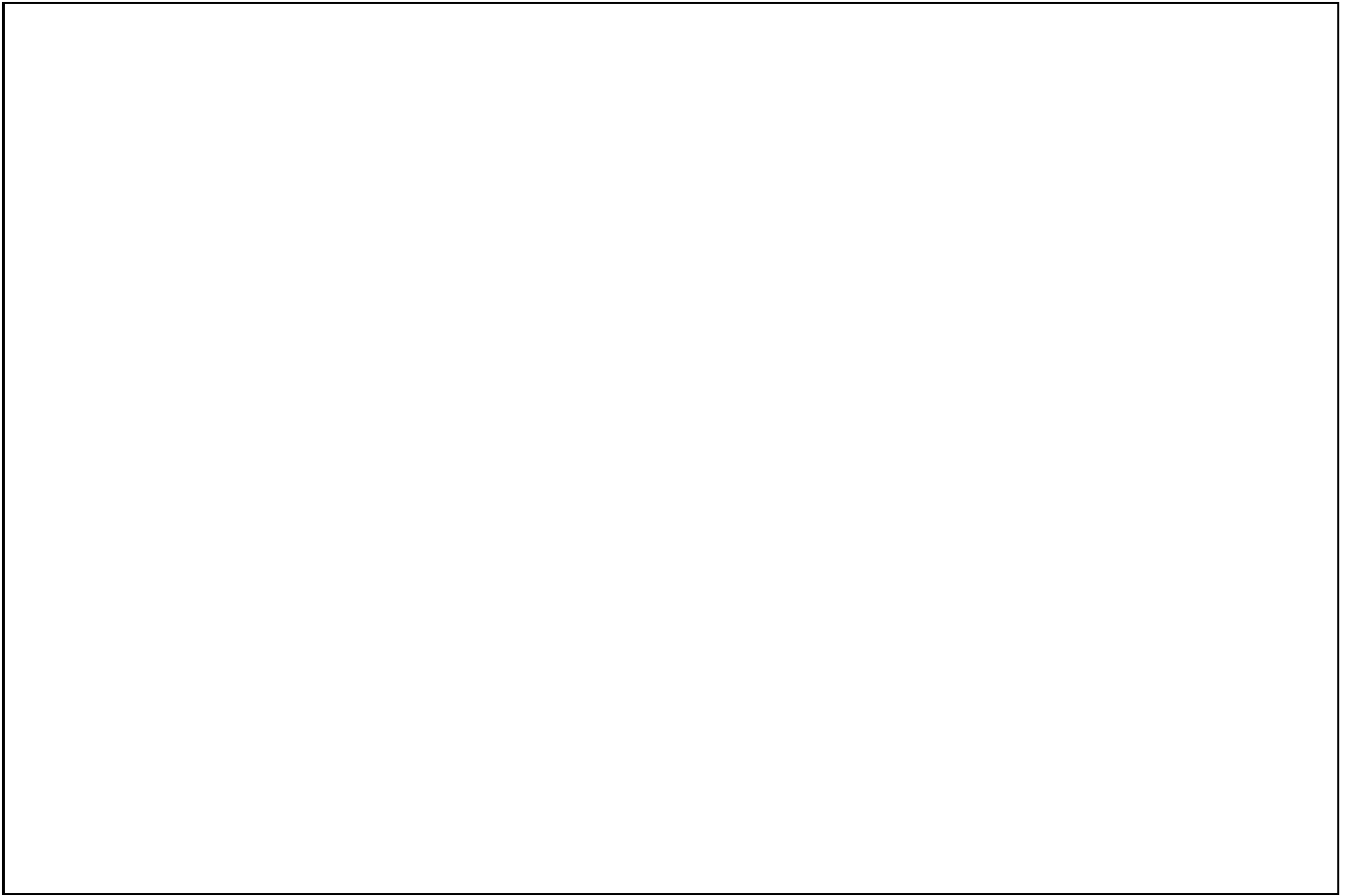
Hertzsprung-Russell diagram of \logg\ vs.~\logt\ with the Kurucz grid
drawn in.  Solid lines are theoretical stellar evolution tracks and
the dashed lines are isochrones for 1, 3, 5, 10, 15, 25 and
$50\Myr$. The ZAMS is indicated with a solid line
}
\endfig

Ages were derived for all the subgroups by fitting  isochrones in  the
\logt, \logg\ plane. This method of age determination suffers from the
small number of stars at the main-sequence turnoff in an association
and the large errors in \logg . Extensive simulations were performed
to investigate the influence of these error sources on the age
determination and in order to choose the best fitting method (appendix
A).  The results show that subgroup 1b is not intermediate in age
between 1a and 1c, as is often assumed, but is in fact the younger of
the three, so that the sequence of increasing age is: 1d, 1b, 1c, 1a.

Visual extinctions derived for all programme stars were compared to
IRAS $100\um$ skyflux data in order to constrain the distances and
depths of the GMCs Orion A and B. We find that the near edge of the
Orion A and B clouds is located at a distance of $\sim 320\pc$ and the
far edge of the clouds is at $\sim 500\pc$.

Simulations show that with the limited amount of data available on the
PDMF of the subgroups in Orion OB1 it is very difficult to constrain
the IMF accurately (appendix B). A stellar IMF of the form $\xi (\log
M)=AM^{-B}$ fits the observations of each of the subgroups 1a, 1b and
1c.  In all three cases the exponent $B$ is $1.7\pm 0.2$. The stellar
content, the ages and the IMF of the subgroups were combined to
calculate the amount of energy that was released into the ISM by Orion
OB1.  These energies are sufficient to account for the Orion-Eridanus
bubble; blow out of this bubble may occur, but this depends critically
on the z-distribution of the surrounding gas.

Orion OB1 displays many indications of sequential star formation. We
speculate that past geometries of the system of gas and stars can
explain our finding that subgroup 1b is younger than 1c. Two possible
scenarios for the origin of the oldest subgroup (1a) were discussed.
The formation of 1a might be the consequence of the impact of a high
velocity cloud in the Galactic disk or it might have formed when the
expanding Lindblad ring fragmented into molecular clouds. At present,
we cannot distinguish between these two alternatives.

Future work includes an analysis of the HIPPARCOS data together with
radial velocities obtained as part of an ESO Key Programme (Hensberge
\ea\ 1990).  This will provide improved membership lists and  possibly
the internal  motions in Orion OB1.  Another interesting project is to
take  a more  detailed look  at  the ISM in the vicinity of Orion OB1.
This can be done by combining the  new \HI data of  Hartmann \& Burton
(1993)  with  existing  CO  and  IRAS  data.  In  this  way  different
components  of the  ISM  can  be studied  and this will allow  a  more
detailed  analysis of the extinctions (see section  4.3), as well as a
more refined analysis  of the  interaction of the  stars  and the ISM.
Work along these lines is in progress.

\acknow {
It is a pleasure to thank A.\ Blaauw and H.\ Lamers for stimulating
discussions.  We thank T.~Chester and S.~Wheelock for providing the
IRAS Sky Survey plates and D.~Kester for help with producing the
mosaic of these plates. We are grateful to the referee, J.\ Hesser,
for a meticulous reading of the manuscript. This research was
supported (in part) by the Netherlands Foundation for Research in
Astronomy (NFRA) with financial aid from the Netherlands organization
for scientific research (NWO).  EdG acknowledges support from NSF
grant AST-8918912 and NASA grant NAG5-1736.  }

\appendix {A: age determination}
The ages  of   the subgroups in    Orion OB1 were derived   by fitting
theoretical isochrones to the observational data in the \logg , \logt\
plane. The age of a subgroup was determined by minimizing:
$$\Psi^2=\sum_{i=1}^N d^2_{{\rm min},i}W_i\eqno (1)\; ,$$
where $d_{{\rm min},i}$ is the shortest distance from the isochrone to
the data point $i$ and $W_i$ is a weight factor, which is different
for each star. This method suffers from a number of problems:
\medskip
\item {$\bullet$} Accidental errors in $\log g$ are large ($\sigma (\log
                  g)\approx 0.25$ for $\log T_{\rm eff} > 4.3$ and
                  $\sigma(\log g) \approx 0.1$ for $\log T_{\rm eff} < 4.3$);
\item {$\bullet$} Due to the steep slope of the IMF there is only a
                  small number of stars at the turnoff of the isochrone;
\item {$\bullet$} Due to the binary or peculiar nature of some stars
                  $\log g$ and $\log T_{\rm eff}$ may be in error;
\item {$\bullet$} Contamination by non-members.
\medskip
We performed Monte Carlo simulations in order to assess how well the
fitting method works, and to investigate the effects of the relatively
large errors in \logg\ and the influence of the slope of the IMF. We
generated synthetic associations by drawing a population of stars from
a distribution which has the same form as the IMF. The association was
aged with the aid of the theoretical evolutionary tracks and the
parameters of the stars in the association were transformed to the
\logg , \logt\ plane. Random (normally distributed) observational
errors were added in accordance with the errors of our observations
(section 3). Figure 11 shows the theoretical evolutionary tracks that
we used. The Kurucz grid is also plotted in this figure to illustrate
that at the high mass end the isochrones cannot be transformed to the
\bminu\ vs.~\bminl\ diagram. Figure 12 shows an example of a synthetic
association in the \logg , \logt\ plane, together with the isochrone
representing its real age.

\begfig 8cm
\figure {12}{
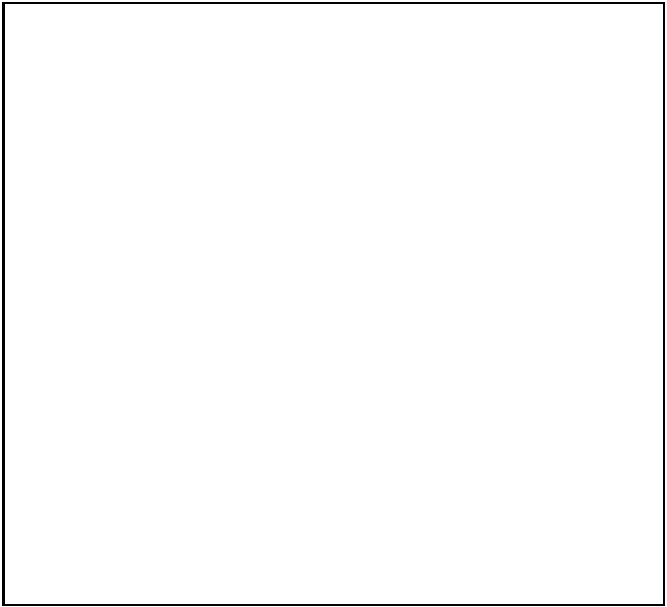
Hertzsprung-Russell diagram of \logg\ vs.~\logt\ for a synthetic
association containing 300 stars with masses between 2 and $120\Msol$.
The association has a single power law IMF: $\xi (\log M)=AM^{-1.6}$
and its age is $5\Myr$. The solid line is the isochrone that
corresponds to the age of the association. Note the influence of the
large errors in \logg\ at high temperatures (the region of the main
sequence turn off)
}
\endfig

The synthetic associations do not contain any binaries; the occurrence
of non-members in the data was also not taken into account. Undetected
binaries may give rise to erroneous points in the \logg , \logt\
plane, as the measured colours are for the composite light rather than
for the individual components. In practice this effect is not very
important. For a large fraction of the binaries we obtained photometry
for both components, and the bulk of the remainder has a companion of
a mass too low to influence the colours. In cases where there are
doubts about the colours of a star because its companion is also in
the diaphragm of the photometer and is bright enough to influence the
colours, the data were simply excluded from the actual fitting
process. Non-members, i.e., interlopers, generally increase the number
of low-mass stars and this often leads to gross overestimates of the
age of an association. To prevent this, stars with $\lt <4.15$ were
excluded from the fitting process. This temperature was chosen because
below this value isochrones for less than $50\Myr$ become practically
indistinguishable to the fitting process (see Fig.~11).

Many realizations of a synthetic association were calculated for
different ages, and we fitted theoretical isochrones for each of them
according to the procedure mentioned above. The results were analyzed
for three effects:
\medskip
\item {$\bullet$} What should the choice be for the weights $W$ in Eq.\
                  (1)?
\item {$\bullet$} What is the influence of the functional form of the IMF?
\item {$\bullet$} What accuracies in \logg\ (and \logt ) are required
                  for accurate age determinations?
\medskip
A weight factor $W$ is introduced in Eq.\ (1) to compensate for the
fact that there are only a few stars at the high-mass end of the IMF
compared to the number of low-mass stars. This means that without
weighting, the stars at the main-sequence turnoff will effectively
play no role in the age determination. Yet, the turnoff contains all
the information. The obvious choice for $W$ would be some power of the
stellar mass, so as to compensate for the shape of the IMF. But
stellar masses contain the errors in \logt ,
\logg\ and $M_{\rm bol}$ and are thus very uncertain.

After some experimentation we found that a good way of assigning
weights is to use $W=\lt$. The errors in \logt\ are relatively small
and it is a physical parameter which follows from the observations,
thus removing the arbitrariness that plagues many other weighting
schemes. Figures 13a and b show the average age determined by
isochrone fitting as a function of the real age, for two different
IMFs (single power law and the Miller \& Scalo form -- see below).
For each age plotted, 1000 simulations were used. The error bars
indicate the rms spreads in the derived ages, and vary between $0.9$
and $2.4\Myr$. There clearly is a smooth correlation between the
estimated and the real ages. The derived ages are systematically
overestimated, but a second degree polynomial can be fitted quite
satisfactorily to the results with a $\chi^2$ of $0.02$ and $0.05$ and
average residuals $0.06$ and $0.1$, for the single power law and the
Miller \& Scalo IMF, respectively. This relation can then be used to
correct ages derived for observed associations.

\begfig 8cm
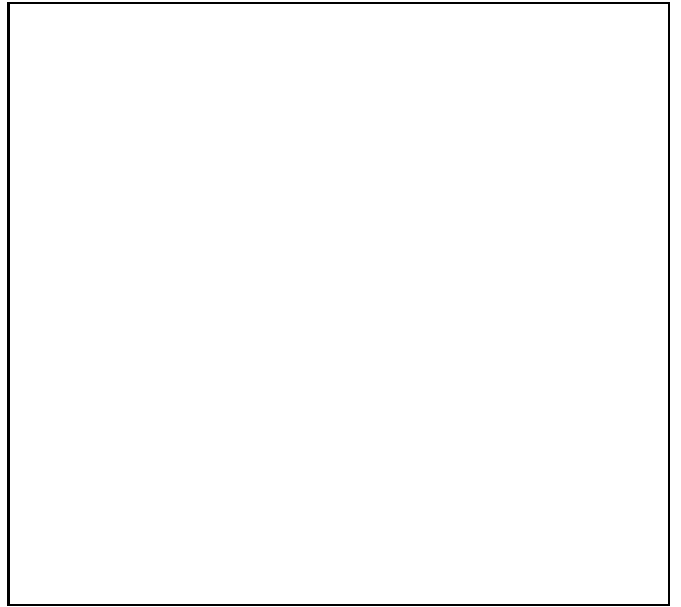
\figure {13a}{
Best estimate of the age (which follows from isochrone fitting)
vs.~the real age of the association. The results for each age are the
average of 1000 simulations. The error bars indicate rms spreads. The
results are shown for a synthetic association that contains 300 stars
between 2 and $120\, \Msol$ and the IMF is: $\xi (\log M)=AM^{-1.6}$
}
\endfig

\begfig 8cm
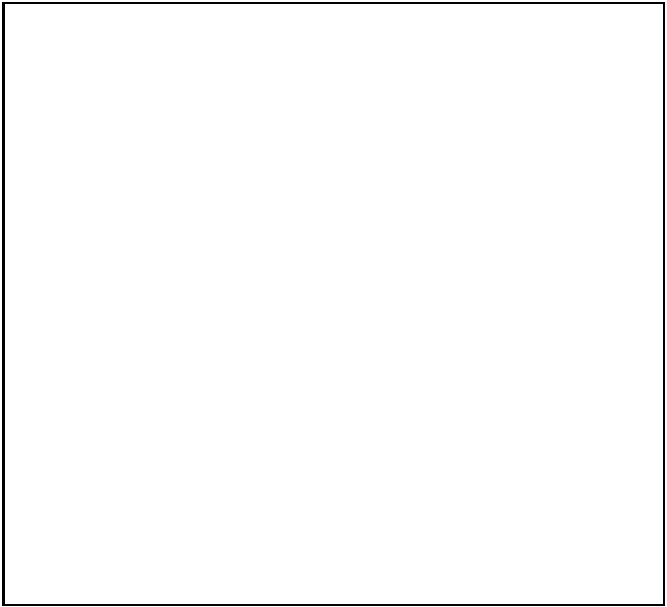
\figure {13b}{
The same as 13a for an association containing the same number of
stars, but following a Miller \& Scalo IMF
}
\endfig

A disadvantage of the \logt\ weighting scheme is that stars at the
turnoff on isochrones of low age can get the same weights assigned
even though their masses are different. This is probably also part of
the cause of the systematic overestimate of the derived ages (the
other part is probably the relatively high number of low mass stars).
In order to check whether this problem can be resolved we also tested
a weighting scheme which is a combination of \logg\ and \logt .  This
would remove the problems with the \logt\ weights at the high mass
end. We tried a scheme with weights $T^4/g$ which is basically $L/M$,
a quantity that increases steeply with mass. This scheme also results
in a smooth trend in derived age vs.~real age, but it is less smooth
than that obtained with the simple \logt\ weights, and higher order
polynomials or more complicated functions are needed to fit the
relation. The ages quoted in section 4.2 were derived with the \logt\
weighting scheme.

The second question that can be answered with the results of the
simulations is whether the shape of the IMF makes any difference in
deriving ages. Again, the most important influence of the IMF is that
there are only a few stars at the main-sequence turnoff for young
associations. Moreover, an IMF with a steeper slope at the high mass
end might make it more difficult to determine correct ages. We tried
two different IMFs, a single power law: $\xi (\log M)=AM^{-1.6}$ (see
e.g., Basu \& Rana 1992); and a Miller \& Scalo IMF: $\xi (\log
M)=AM^{-B}$, where $B=0.4$ for $M<1\Msol$, $B=1.5$ for $1<M<10\,
\Msol$ and $B=2.1$ for $M>10\Msol$ (see De Geus 1992). The results for
the two IMFs look similar as shown in Figs.~13a and b, and the derived
ages are all consistent within the rms spreads, although for ages
below $\sim 5\Myr$ the single power law IMF results in somewhat
smaller overestimates. Also, when correcting the derived age to the
real age, the results for the two IMFs are entirely consistent within
the error bars.

The fact that the difference in the IMFs cannot be seen in the results
implies that the age determinations are dominated by the observational
errors in \logg\ and \logt. The simulations were repeated with lower
observational errors and it turns out that if the observational
errors, and in particular those in \logg , were a factor of three
smaller, then the ages can be derived with an accuracy of $\pm
0.5\Myr$. This translates to requiring that bolometric magnitudes be
determined with errors less than $0.2$--$0.4$ magnitudes.
Improvements in the absolute magnitude calibration can be expected
when HIPPARCOS data become available for nearby early-type stars.

\appendix{B: the initial mass function}
There are a number of ways to determine the stellar IMF for a group of
stars. The conventional approach is to plot the number of stars as a
function of mass in a $\log$-$\log$ diagram and fit a power law IMF to
the data. But if the star formation process is viewed as drawing a
population of stars randomly from a distribution function of the same
form as the IMF (see section 4.4), the conventional approach is not
the right method. In such a procedure, fitting a power law to the
observations does not take into account that a particular population
of stars drawn from the IMF does not necessarily reflect the
underlying distribution of masses accurately.

\begfig 8cm
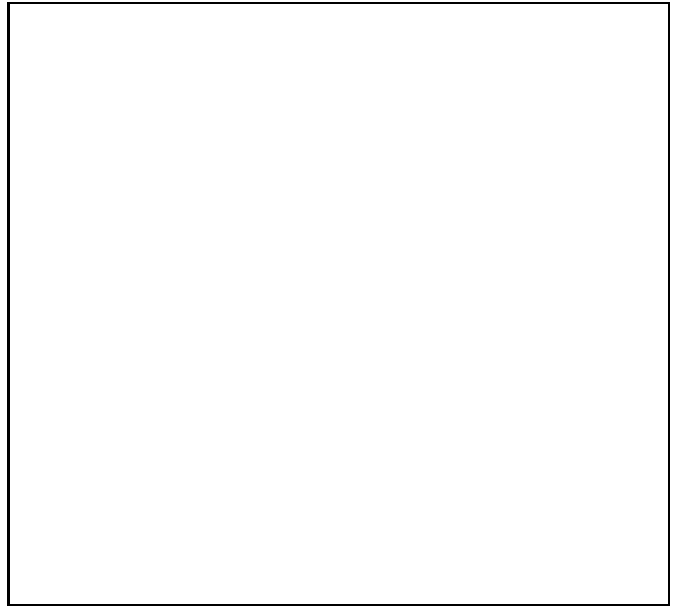
\figure {14a}{
Mean (from 1000 simulations) of the probability (P) that the synthetic
association, generated according to a Miller \& Scalo IMF, has the
single power law IMF that is indicated on the abscissa (by its power
law index). The square indicates that the mass distribution was
compared to a Miller \& Scalo IMF. Parameters of the synthetic
association are indicated
}
\endfig

\begfig 8cm
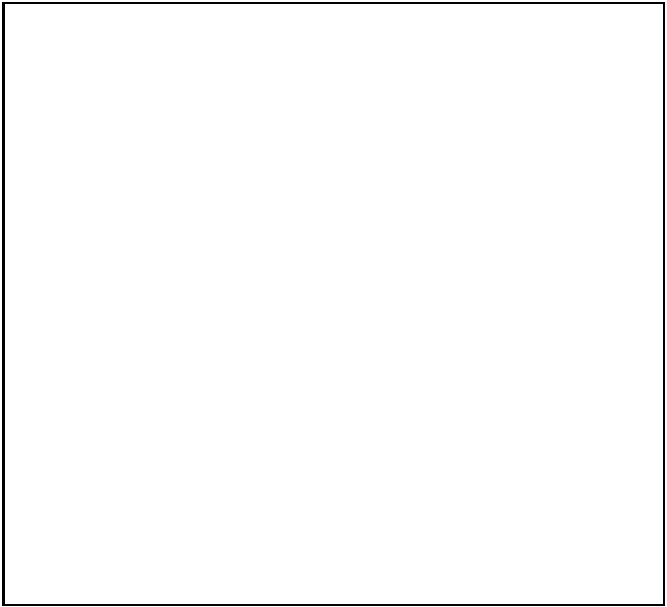
\figure {14b}{
Same as 14a for a synthetic association generated according to a
single power law IMF
}
\endfig

\begfig 8cm
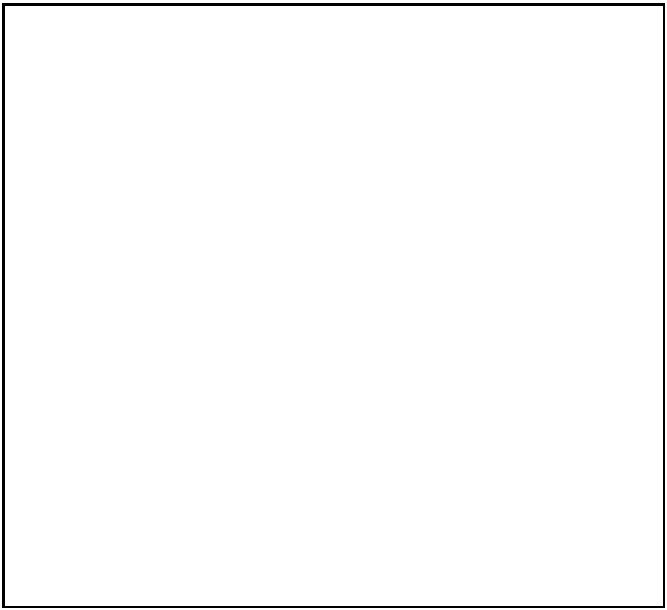
\figure {15a}{
The same as Fig.~14, but now the results obtained when applying the
KS-test to the actual present day mass function of subgroup 1a are
shown. The parameters of the PDMF are listed in Table 4
}
\endfig

\begfig 8cm
\figure {15b}{
Same as 15a for subgroup 1b
}
\endfig

\begfig 8cm
\figure {15c}{
Same as 15a for subgroup 1c
}
\endfig

The statistical test that can be applied to check whether an observed
population could have been drawn form a certain distribution is the
Kolmogorov-Smirnoff test. The procedure in this case is to determine
first the present day mass function (PDMF), and compare it to assumed
forms of the IMF with the aid of the KS-test. However,
\medskip
\item {$\bullet$} The PDMF is known completely only in a limited mass
                  interval, with the limits being set by the minimum mass at
                  which the observations are complete and by the maximum mass
                  which follows from the age of the subgroup;
\item {$\bullet$} The number of stars in this interval is limited: 20
                  to 50 stars in the case of the Orion OB1 subgroups;
\item {$\bullet$} For high mass stars ($M\ga 10\Msol$) there are large
                  uncertainties in the masses, caused by the large
                  errors in \logg\ (see also section 3).
\medskip
We performed Monte Carlo simulations in order to investigate how well
the KS-test can distinguish between different forms of the IMF. As in
Appendix A, this was done by generating synthetic associations
according to the IMF and applying the KS-test on a limited mass
interval (4--$30\Msol$) with a number of stars similar to what is seen
in Orion OB1. The PDMF of the synthetic association was compared to a
range of single power law IMFs (indices varying from $-0.4$ to $-2.1$)
and to the Miller \& Scalo IMF (see Appendix A). The synthetic
associations had a single power law IMF (index $-1.6$) or the Miller
\& Scalo IMF. The results are shown in Fig.~14 where the mean of the
probability that the PDMF is drawn from the IMF is plotted against the
form of the comparison IMF. We used 1000 realizations of the synthetic
association.  We conclude:
\medskip
\item {$\bullet$} Based on the small number of observations it is
                  nearly impossible to distinguish between different single
                  power law IMFs (although the diagrams might look better in
                  individual cases).
\item {$\bullet$} A single power law IMF with an index close to $-1.5$
                  cannot be distinguished from a Miller \& Scalo IMF and vice
                  versa, which is partly due to the limited mass range
		  over which we fit the PDMF to the IMF.
\medskip
{}From Fig.~14 it is evident that the error bars on the power law index
of the IMF are rather large. In fact, if one were to be conservative
and demand a confidence level of 95\% for the derived index, then
almost every IMF is consistent with the data. Again, a better
determination may be possible in individual cases. The simulations
showed that the error bars on the index are much smaller if the PDMF
contains the data of $\sim 100$ stars or more.

Figure 15 shows the results for the Orion OB1 subgroups. For subgroup
1a, the IMF that is most consistent with the observations is the
single power law IMF with power law index $-1.7$, but, demanding a
confidence level of 70\% , all single power law IMFs with indices
between $-1.3$ and $-2.1$ and the Miller \& Scalo IMF are consistent
with the observations. Similarly, for subgroup 1b the index is $-1.6$,
with indices $-1.2$ to $-1.9$ and the Miller \& Scalo IMF being
consistent at the 70\% confidence level. In the case of subgroup 1c
(note the effect of the very low number of stars), $-1.7$ is the most
consistent index, with indices between $-1.0$ and $-2.1$ and the
Miller \& Scalo IMF being consistent with the data at the 70\%
confidence level. Taking into account the large error bars, we adopt a
single power law IMF with index $-1.7$ for all three subgroups.

\begref {References}
\ref
van Altena, W.F., Lee, J.T., Lee, J.-F., Lu, P.K., Upgren, A.R., 1988,
\aj 95, 1744

\ref
Anthony-Twarog, B.J., 1982, \aj 87, 1213

\ref
Bally, J., Langer, W.D., Stark, A.A., Wilson, R.W., 1987, \apj 312,
L45

\ref
Bally, J., Langer, W.D., Wilson, R.W., Stark, A.A., Pound, M.W., 1991,
Fragmentation of Molecular Clouds and Star Formation , IAU Symp.\ 147,
eds.\ E.\ Falgarone, F.\ Boulanger \& G.\ Duvert (Dordrecht: Kluwer), p.\ 11

\ref
Bastian, U., R\" oser, S., Nesterov, V.V. \ea , 1991, \aapss 87, 159

\ref
Basu, S., Rana, N.C., 1992, \apj 393, 373

\ref
Beichman, C.A., 1987, \araa 25, 521

\ref
Blaauw, A., 1961, Bull.\ Astr.\ Inst.\ Netherlands 15, 265

\ref
Blaauw, A., 1964, \araa 2, 213

\ref
Blaauw, A., 1991, in The Physics of Star Formation and Early Stellar
Evolution, eds.\ C.J. Lada \& N.D.\ Kylafis, NATO ASI Series C,
Vol.\ 342, p.\ 125

\ref
Blaauw, A., 1993, in Massive Stars: Their Lives in The Interstellar
Medium, eds. J.P.\ Casinelli \& E.B.\ Churchwell, ASP conference series,
Vol.\ 35, p.\ 207

\ref
Brand, J., Wouterloot, J.G.A., 1988, \aapss 75, 117

\ref
Burrows, D.N., Singh, K.P., Nousek, J.A., Garmire, G.P., Good, J.,
1993, \apj 406, 97

\ref
Castor, J., McCray, R., Weaver, R., 1975, \apj 200, L107

\ref
Cayrel, R., 1990, in Physical Processes in Fragmentation and Star
Formation, eds.\ R.\ Capuzzo-Dolcetta, C.\ Chiosi \& A.\ Di Fazio, p.\ 343

\ref
Claudius, M., Grosb\o l, P.J., 1980, \aap 87, 339

\ref
Chromey, F.R., Elmegreen, B.G., Elmegreen, D.M., 1989, \aj 98, 2203

\ref
Comer\' on, F., Torra, J., 1992, \aap 261, 94

\ref
Cowie, L.L., Songaila, A., York, D.G., 1979, \apj 230, 469

\ref
Crawford, D.L., 1978, \aj 83, 48

\ref
Crawford, D.L., Barnes, J.V., 1966, \aj 71, 610

\ref
Elmegreen, B.G., 1992, in Star Formation in Stellar Systems,
eds.\ G.\ Tenorio-Tagle, M.\ Prieto \& F.\ S\' anchez, Cambridge
University Press, p.\ 383

\ref
Elmegreen, B.G., Lada, C.J., 1977, \apj 214, 725

\ref
Franco, J., Tenorio-Tagle, G., Bodenheimer, P., R\' o\. zycka, M.,
Mirabel, I.F., 1988, \apj 333, 826

\ref
de Geus, E.J., 1991, in The Formation and Evolution of Star Clusters,
ed.\ K.\ Janes, ASP conference series, Vol.\ 13, p.\ 40

\ref
de Geus, E.J., 1992, \aap 262, 258

\ref
de Geus, E.J., Burton, W.B., 1991, \aap 246, 559

\ref
de Geus, E.J., de Zeeuw, P.T., Lub, J. (GZL), 1989, \aap 216, 44

\ref
de Geus, E.J., Lub, J., van de Grift, E., 1990, \aapss 85, 915

\ref
Genzel, R., Stutzki, J., 1989, \araa 27, 41

\ref
Gies, D.R., Bolton, C.T., 1986, \apjss 61, 419

\ref
Groe\-ne\-we\-gen, M.A.T., La\-mers, H.J.G.L.M., Paul\-drach, A.\ W.A.,
1989, \aap 221, 78

\ref
Goudis, C., 1982, The Orion Complex: A Case Study of Interstellar
Matter, Astrophys.\ Space Sci.\ Libr., Vol.\ 90, Dordrecht: Reidel

\ref
Green, D.A., 1991, \mn 253, 350

\ref
Green, D.A., Padman, R., 1993, \mn 263, 535

\ref
Hardie, R.H., Heiser, A.M., Tolbert, C.R., 1964, \apj 140, 1472

\ref
Hartmann, D., Burton, W.B., 1994, in preparation

\ref
Heiles, C., 1976, \apj 208, L137

\ref
Hensberge, H., van Dessel, E.L., Burger, M. \ea , 1990, The Messenger,
61, 20

\ref
Hoogeveen, S.J., 1992, \apss 196, 299

\ref
Jones, B.F., Walker, M.F., 1988, \aj 95, 1755

\ref Jung, J., Bischoff, M., 1971, Bulletin d'Information du
Centre de Donnees Stellaires 2, 8

\ref
Kudritzki, R.P., Pauldrach, A., Puls, J., Abbott, D.C., 1989,
\aap 219, 205

\ref
Kurucz, R.L., 1979, \apjss 40, 1

\ref
Kutner, M.L., Tucker, K.D., Chin, G., Thaddeus, P., 1977, \apj 215,
521

\ref
Lada, E.A., Bally, J., Stark, A.A., 1991, \apj 368, 432

\ref
Lamers, H.J.G.L.M., Leitherer, C., 1993, \apj 412, 771

\ref
Lesh, J.R., 1968, \apj 152, 905

\ref
Leitherer, C., Robert, C., Drissen, L., 1992, \apj 401, 596

\ref
Leonard, P.J.T., Duncan, M.J., 1988, \aj 96, 222

\ref
Leonard, P.J.T., Duncan, M.J., 1990, \aj 99, 608

\ref
Lindblad, P.O., Grape, K., Sandqvist, Aa., Schober, J., 1973,
\aap 24, 309

\ref
Lub, J., Pel, J.W., 1977, \aap 54, 137

\ref
Lockman, F.J., Hobbs, L.M., Shull, J.M., 1986, \apj 301, 380

\ref
Mac Low, M.-M., McCray, R., 1988, \apj 324, 776

\ref
Maddalena, R.J., Morris, M., Moscowitz, J., Thaddeus, P., 1986,
\apj 303, 375

\ref
Maeder, A., 1972, in IAU Colloquium No.\ 17, Stellar Ages,
eds.\ G.\ Cayrel de Strobel \& A.M.\ Delplace (Meudon: Observatoire de
Paris), \S XXIV

\ref
McCray, R., Kafatos, M.C., 1987, \apj 317, 190

\ref
Miller, G.E., Scalo, J.M., 1979, \apjss 41, 513

\ref
Morgan, W.W., Lod\' en, K., 1966, Vistas in Astronomy 8, 83

\ref
Morrell, N., Levato, H., 1991, \apjss 75, 965

\ref
Olano, C.A., 1982, \aap, 112, 195

\ref
Parenago, P.P., 1953, AZh. 30, 249 (Eng\-lish Trans.\ in
Astr.\ News\-letter 74, 20)

\ref
Parenago, P.P., 1954, Publ.\ Sternberg Astron.\ Inst.\ No.\ 25

\ref
Reynolds, R.J., Ogden, P.M., 1979, \apj 229, 942

\ref
R\" oser, S., Bastian, U., 1989, PPM-postions and proper motions of
181731 stars north of $-2.5$ degrees declination,
Astron.\ Rechen-Inst.\ Heidelberg.

\ref
Rowan-Robinson, M., Hughes, J., Jones, M. \ea , 1991, \mn 249, 729

\ref
Schaller, G., Schaerer, D., Meynet, G., Maeder, A., 1992,
\aapss, 96, 269

\ref
Sharpless, S., 1952, \apj 116, 251

\ref
Sharpless, S., 1962, \apj 136, 767

\ref
Shull, J.M., 1993, in Massive Stars: Their Lives in the Interstellar
Medium, eds.\ J.P.\ Casinelli \& E.B.\ Churchwell, ASP conference series
Vol.\ 35, p.\ 327

\ref
Sivan, J.P., 1974, \aapss 16, 163

\ref
Smart, R.L., 1993, PhD.\ thesis University of Florida

\ref
Strai\v zys, V., Kuriliene, G., 1981, \apss 80, 353

\ref
Strand, K.Aa., 1958, \apj 128, 14

\ref
Tenorio-Tagle, G., Bodenheimer, P., 1988, \araa 26, 145

\ref
Tomisaka, K., Ikeuchi, S., 1986, \pasj 38, 697

\ref
Turon, C., Cr\' ez\' e, M., Egret, D. \ea , 1992, ESA SP-1136

\ref
Vilkoviskij, E.,Ya., Tambovtseva, L.V., 1992, \aapss 94, 109

\ref
Walker, M.F., 1969, \apj 155, 447

\ref
Warren, W.H., Hesser, J.E., 1977 (WH), \apjss 34, 115

\ref
Warren, W.H., Hesser, J.E., 1978 (WH), \apjss 36, 497

\ref
Weaver, R., McCray, R., Castor, J., Shapiro, P., Moore, R., 1977, \apj
218, 377 (err 220, 742)

\ref
Wheelock, S. \ea, 1993, IRAS Sky Survey Atlas Explanatory Supplement

\ref
de Zeeuw, P.T., Brand, J., 1985, in Birth and Evolution of Massive
Stars and Stellar Groups, eds.\ W.\ Boland \& H.\ van Woerden, Dordrecht:
Reidel, p.\ 95

\bye